\documentclass[sigconf]{acmart}

\usepackage{listings}
\usepackage{caption}
\usepackage{subcaption}
\usepackage{multirow}
\usepackage{enumitem}
\usepackage{stackengine}
\usepackage[frozencache]{minted}
\newsavebox{\fmbox}
\newenvironment{smpage}[1]
{\begin{lrbox}{\fmbox}\begin{minipage}{#1}}
{\end{minipage}\end{lrbox}\usebox{\fmbox}}

\usepackage{tikz}

\copyrightyear{2021}
\acmYear{2021}
\setcopyright{acmlicensed}\acmConference[CHI '21]{CHI Conference on Human Factors in Computing Systems}{May 8--13, 2021}{Yokohama, Japan}
\acmBooktitle{CHI Conference on Human Factors in Computing Systems (CHI '21), May 8--13, 2021, Yokohama, Japan}
\acmPrice{15.00}
\acmDOI{10.1145/3411764.3445249}
\acmISBN{978-1-4503-8096-6/21/05}

\begin{document}

\title{Falx: Synthesis-Powered Visualization Authoring}

\author{Chenglong Wang}
\email{clwang@cs.washington.edu}
\affiliation{%
  \institution{University of Washington}
}

\author{Yu Feng}
\email{yufeng@cs.ucsb.edu}
\affiliation{%
  \institution{University of California, Santa Barbara}
}

\author{Rastislav Bodik}
\email{bodik@cs.washington.edu}
\affiliation{%
  \institution{University of Washington}
}

\author{Isil Dillig}
\email{isil@cs.utexas.edu}
\affiliation{%
  \institution{The University of Texas at Austin}
}

\author{Alvin Cheung}
\email{akcheung@cs.berkeley.edu}
\affiliation{%
  \institution{University of California, Berkeley}
}

\author{Amy J. Ko}
\email{ajko@uw.edu}
\affiliation{%
  \institution{University of Washington}
}

\renewcommand{\shortauthors}{Chenglong Wang et al.}

\begin{abstract}
Modern visualization tools aim to allow data analysts to easily create exploratory visualizations. When the input data layout conforms to the visualization design, users can easily specify visualizations by mapping data columns to visual channels of the design. However, when there is a mismatch between data layout and the design, users need to spend significant effort on data transformation. 

We propose Falx, a synthesis-powered visualization tool that allows users to specify visualizations in a similarly simple way but without needing to worry about data layout. In Falx, users specify visualizations using examples of how concrete values in the input are mapped to visual channels, and Falx automatically infers the visualization specification and transforms the data to  match the design. In a study with 33 data analysts on four visualization tasks involving data transformation, we found that users can effectively adopt Falx to create visualizations they otherwise cannot implement.
\end{abstract}

\maketitle
\thispagestyle{empty}

\section{Introduction}

\begin{figure}[h]
\small
\centering
     \begin{subfigure}[c]{0.15\textwidth}
         \centering
         \begin{tabular}{|c|c|c|}
			\hline
			\textsf{Date} & \textsf{Temp} & \textsf{Type}\\\hline
			09-05 & 64.4 & Low\\
			09-05 & 87.8 & High\\
			09-06 & 53.6 & Low\\
			09-06 & 80.6 & High\\\hline
			\end{tabular}
     \end{subfigure}
     \quad$\xrightarrow{~
    		    \stackon{
                    \stackon
                        {\small $\textsf{Type}\rightarrow \mathit{color}$}
                        {\small $\textsf{Temp}\rightarrow y$}}
    		       {\small $\textsf{Date}\rightarrow x$}~}$
    ~~
     \begin{subfigure}[c]{0.18\textwidth}
         \centering
		\includegraphics[width=1.1\textwidth]{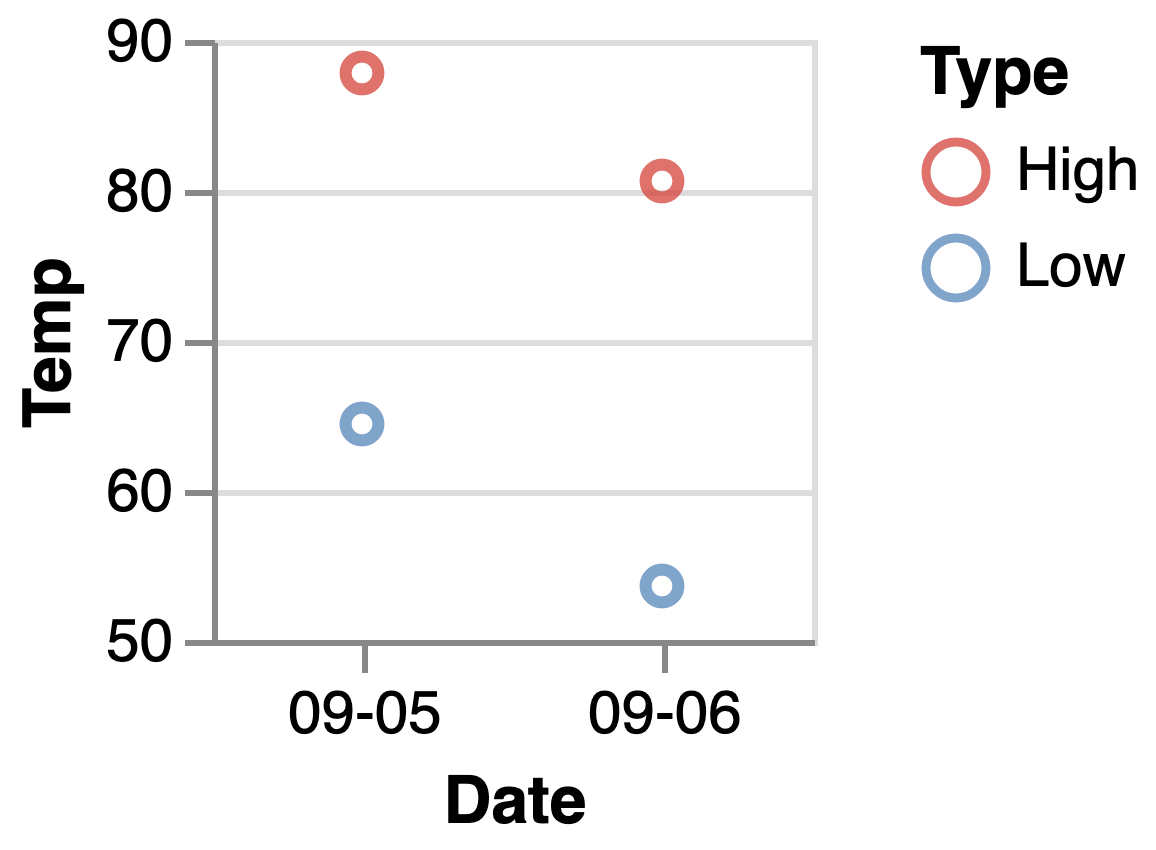}
     \end{subfigure}
\vspace{-10pt}
\caption{An example dataset and its scatter plot visualization that maps {\normalfont \textsf{Date}} to $x$, {\normalfont \textsf{Temp}} to $y$ and {\normalfont \textsf{Type}} to $\mathit{color}$.}
\label{fig:simple-vis-example}
\end{figure}

Modern visualization authoring tools, such as declarative visualization grammars like ggplot2~\cite{wickham2011ggplot2}, Vega-Lite~\cite{DBLP:journals/tvcg/SatyanarayanMWH17} and interactive visualization tools like Tableau~\cite{DBLP:conf/kdd/StolteTH02} and Voyager~\cite{wongsuphasawat2015voyager}, are built to reduce data analysts' efforts in authoring visualizations in exploratory data analysis. At the heart of these  tools, visualizations are specified using grammars of graphics~\cite{wilkinson2012grammar}, where every visualization can be succinctly specified using the following three components:
\begin{itemize}[leftmargin=5.5mm]
\item A graphical mark that defines the geometric objects used to visualize the data (e.g., line, scatter plots, bars),
\item A set of visual encodings that map data variables to visual channels (e.g., $x$, $y$-positions of points),
\item A set of parameters that decide visualization details: coordinate system, scales of axes, legends and titles.
\end{itemize}
In practice, users only need to specify the mark and the visual encodings in order to create the visualization because many tools use a rule-based engine to automatically fill in parameters for visualization details (often referred to as ``smart defaults'') unless the user wants further customization. The abstraction of graphical marks, visual encoding channels, and adoption of smart default parameters open an expressive design space for data analysts that allow them to rapidly construct visualizations for exploratory analysis through simple specifications~\footnote{In our paper, we refer to ``expressive visualizations'' as the set of visualizations that are supported by tools powered by grammars of graphics (e.g., visualizations in Tableau, Vega-Lite, ggplot2) as opposed to more general customized visualizations.}. For example, to visualize the dataset in \autoref{fig:simple-vis-example} with three columns \textsf{Date}, \textsf{Temp} (for temperature) and \textsf{Type} as a scatter plot, the user can choose the graphical mark ``point'' with encodings $\{x\mapsto \textsf{Date}, y\mapsto \textsf{Temp}, \mathit{color}\mapsto \textsf{Type}\}$. The visualization tool then creates one point for each row in the input data, by mapping its values in columns \textsf{Date} and \textsf{Temp} to $x$,$y$-positions and assigning a color to each point based on its value in column \textsf{Type}. Here, the tool uses the default linear scale for $x$,$y$-axis and categorical scale for color, which are default parameters that the user does not need to specify explicitly. The final visualization is rendered in \autoref{fig:simple-vis-example} (right).

\begin{figure*}[h]
\small
\centering
\begin{smpage}{\linewidth}
    \centering
    \begin{tabular}{|c|c|c|}
        \hline
        \textsf{Date} & \textsf{Temp} & \textsf{Type}\\\hline
        09-05 & 64.4 & Low\\
        09-05 & 87.8 & High\\
        09-06 & 53.6 & Low\\
        09-06 & 80.6 & High\\\hline
    \end{tabular}
    \quad$\xrightarrow{~\textsf{pivot}~}{}$\quad
    \begin{tabular}{|c|c|c|}
        \hline
        \textsf{Date} & \textsf{Low} & \textsf{High}\\\hline
        09-05 & 64.4 & 87.8\\
        09-06 & 53.6 & 80.6\\\hline
        \multicolumn{3}{c}{}\\
    \end{tabular}
    \quad$\xrightarrow{~
            \stackon{
                \stackon
                    {\small $\textsf{Low}\rightarrow y_\mathit{min}$}
                    {\small $\textsf{High}\rightarrow y_\mathit{max}$}}
               {\small $\textsf{Date}\rightarrow x$}~}$
    \quad
    \begin{smpage}{0.15\textwidth}
        \includegraphics[width=\textwidth]{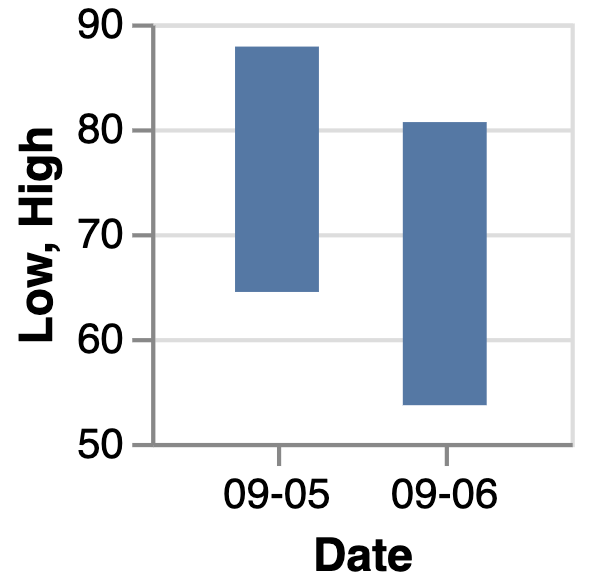}
    \end{smpage}
\end{smpage}
\vspace{-10pt}
\caption{A different visualization design requires transformation of the original input data.}
\label{fig:simple-vis-example-2}
\end{figure*}

In fact, the simplicity of these high-level visualization grammars is grounded in their abstract data model. These grammars expect that the input table is organized in a layout that matches the visualization design~\cite{wickham2014tidy}: (1) each relation forms a row in the input data and corresponds to exactly one geometric object in the visualization, and (2) each data variable forms a column that can be mapped to a visual channel. In practice, however, the mismatch between the data layout and the visualization design is common due to the following reasons~\cite{gatto2015making, wickham2014tidy}:
\begin{itemize}[leftmargin=5.5mm]
\item Tables exported from different sources (e.g., database, analysis tool, different team member) may have different layouts and they may not directly match the visualization design.
\item Different analysis tasks require different visualization designs, and changes in the design can lead to different expected data layout. 
\item The data may need aggregation (e.g., average, count, culminative sum) or additional computation to derive new values prior to visualization.
\end{itemize}
In all of these cases, data analysts cannot directly visualize the data with a simple specification. They have to conceptualize the expected data layout and utilize data transformation tools (e.g., tidyverse~\cite{wickham2014tidy}, Trifacta~\cite{kandel2011wrangler}) to transform the data to match the visualization design. These additional tasks create a barrier for data visualizations and greatly increase the effort required for exploratory analysis~\cite{gatto2015making,morpheus,wongsuphasawat2019goals,2012-enterprise-analysis-interviews}. For example, if the data analyst decides to change the visualization in \autoref{fig:simple-vis-example} to a bar chart with floating bars that show the temperature range  during each day (\autoref{fig:simple-vis-example-2} right), the original data layout will no longer match the new design since the new design expects three data columns (date, lowest temperature, highest temperature) that map to $x$, $y_\mathit{max}$ and $y_\mathit{min}$. As a result, the data analyst needs to transpose the table in \autoref{fig:simple-vis-example}  using a pivot operation (to collect key-values pairs in the \textsf{Type} and \textsf{Temp} columns into new columns) before mapping data columns to  visual channels (\autoref{fig:simple-vis-example-2} right).

We propose Falx, a synthesis-based visualization authoring tool to address the challenges outlined above.~\footnote{Demo available at \url{https://falx.cs.washington.edu/}} Falx builds on recent advances in program synthesis: many program synthesis tools (e.g., FlashFill~\cite{DBLP:conf/popl/Gulwani11}, Wrex~\cite{drosos2020wrex}) have been developed with the promises of automating challenging or repetitive programming tasks for end users by synthesizing programs from user demonstrations. In our design, instead of asking analysts to transform data and specify visualization manually, Falx asks analysts to {\em demonstrate} the visualization task using examples of mappings from concrete values in the input data (as opposed to table columns) to visual channels. Using these examples, Falx automatically synthesizes the programs to transform and visualize the full data, such that resulting visualizations are consistent with the examples (i.e., all example mappings are contained within the visualization). For  example, for the data in \autoref{fig:simple-vis-example-2}, the user can create an example bar \begin{smpage}{0.5\linewidth}\includegraphics[width=\textwidth]{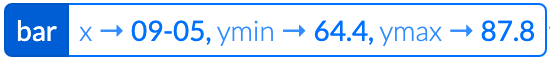}\end{smpage} to demonstrate the task and let Falx create the desired visualization for the full dataset (\autoref{fig:simple-vis-example-2} right). Sometimes, the examples can be ambiguous to Falx, and Falx may generate multiple visualizations that match the example but not necessarily the user intent. In such cases, analysts can interact with an exploration panel to inspect the synthesized visualizations and select the desired one. After that, analysts can further fine-tune details of the desired visualization through a post-processing panel. 

Falx's design has many potential advantages. First, users of Falx specify visualizations by mapping values to visual channels: this approach inherits the simplicity from grammars of graphics but provides more expressiveness since users can use the same examples to specify visualization ideas for inputs with different layouts. 
Second, Falx offloads the data transformation task to the program synthesizer so that users no longer need to conceptualize the expected data layout or transform the data. Finally, while program synthesizers by design can generate multiple results, users can effectively select and validate the desired visualization from synthesized candidates using the exploration panel in Falx. In general, rather than having to construct a visualization, data analysts demonstrate the task using examples and then select the desired visualization from a candidate pool, which shifts from the challenges of expression to the ease of recognition. With these designs, Falx aims to eliminate users’ prerequisites in data transformation and enable data analysts to rapidly author expressive visualizations.

We conducted a user study with 33 participants to test these design hypotheses, studying how users adapt to the new visualization process. Our results show that users of Falx, regardless of previous experience in visualization, can efficiently learn and solve challenging visualizations tasks that cannot be easily solved using the baseline tool ggplot2. However, we also discovered challenges that users face when using the tool and strategies they adopt to solve the problems. We believe these discoveries lead to future opportunities in adopting synthesized-based visualization tools in practice and unveil other potential designs that can further improve the usability of such tools.

\section{Usage Scenario}~\label{sec:usage-scenario}
We first go through an example to illustrate the anticipated user experience in Falx (\autoref{sec:falx-user-experience}) compared to R (\autoref{sec:r-user-experience}). In this example, a data analyst has the following dataset with New York and San Francisco temperature records from 2011-10-01 to 2012-09-30. 

\begin{center}
\medskip
\small
\centering
\begin{tabular}{|c|c|c|}
\hline
Date & New York & San Francisco\\\hline
2011-10-01 & 63.4 & 62.7\\
2011-10-05 & 64.2 & 58.7\\
... & ... & ... \\
2012-09-25 & 63.2 & 53.3\\
2012-09-30 & 62.3 & 55.1\\\hline
\end{tabular}
\medskip
\end{center}

\begin{figure}[ht]
\centering
\includegraphics[width=\linewidth]{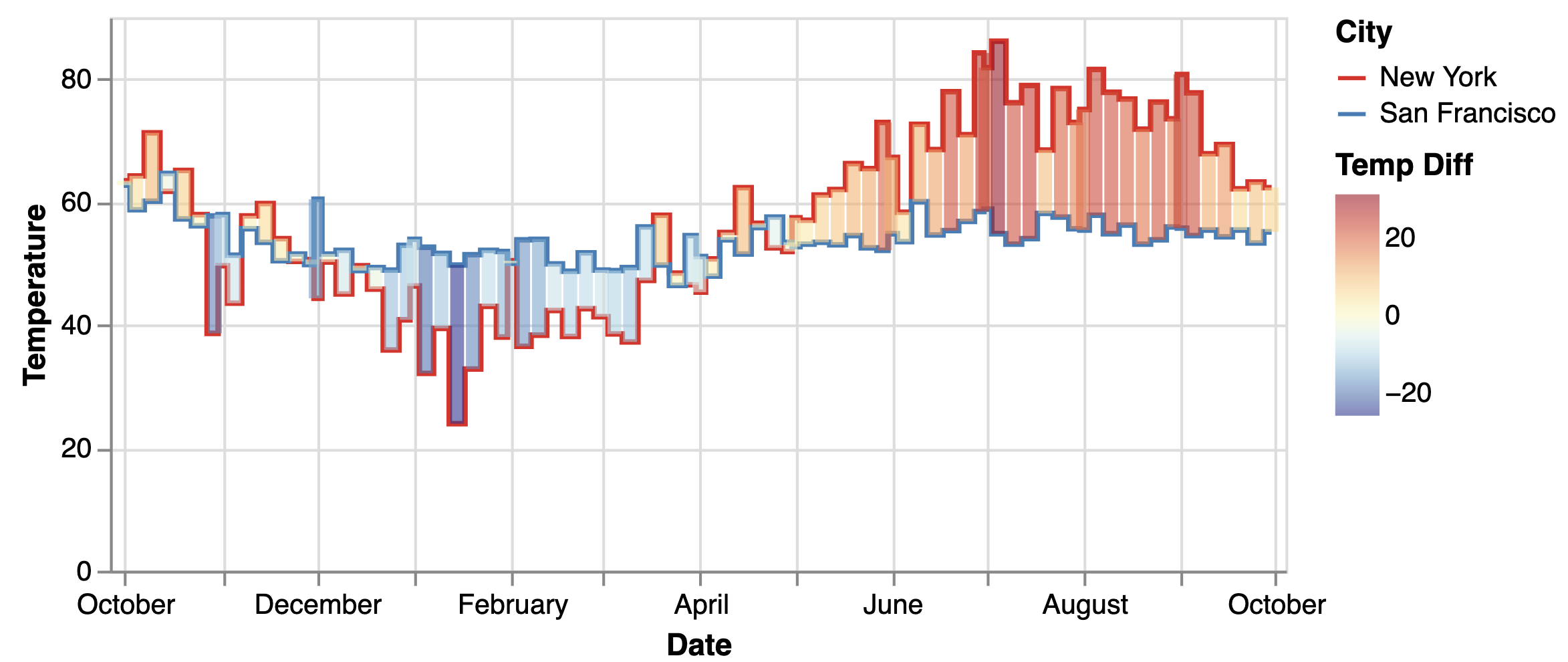}
\caption{A visualization that compares New York and San Francisco temperatures between 2011-10-01 and 2012-09-30.}
\label{fig:nyc-sf-temp-diff}
\end{figure}

\noindent The analyst wants to create a visualization to compare the temperature in the two cities. First, the visualization should contain two lines to show temperature trends in the two cities; these two lines should be distinguished by color. Second, on top of the line chart, a bar chart should be layered on top to show the temperature difference between the two cities for each date. Each bar should start from the New York temperature and end at the corresponding San Francisco temperature, and the color gradient of the bar should indicate the temperature difference between the two cities on that day. The desired visualization is shown in \autoref{fig:nyc-sf-temp-diff}. 

\subsection{User Experience in R}
\label{sec:r-user-experience}

We first illustrate how a data analyst, Eunice, would create this visualization in R using tidyverse~\cite{wickham2014tidy} and ggplot2~\cite{wickham2011ggplot2}, two widely-used libraries for data transformation and data visualization.

After loading the data into a data frame in R, Eunice decides  to first create the line chart that shows temperature trends of the two cities. To do so, Eunice chooses the function \verb|geom_line| from the ggplot2 library. In order to create lines with different colors for different categories, Eunice needs to supply four data variables to the \verb|geom_line| function -- two variables for specifying x and y positions, one for colors of the line, and the last one for groups of lines (i.e., which points belong to the same line). Since the input data does not have these variables, Eunice needs to use the tidyverse library to transform the input data. To do so, Eunice first conceptualizes the desired data layout: the data should have 3 fields---date (for $x$-axis), temperature (for $y$-axis), and city name (for color and group). Eunice recalls a function \verb|pivot_longer| in tidyverse, which supports pivoting the table from a ``wide'' to a ``long'' format by collecting column names and values in the column as key-value pairs in the body content. Specifically, Eunice writes the following code to transform the data, which yields the data on the right that matches Eunice's expectation.

\begin{figure}
\begin{subfigure}{0.98\linewidth}
 \small
 \centering
 \includegraphics[width=\textwidth]{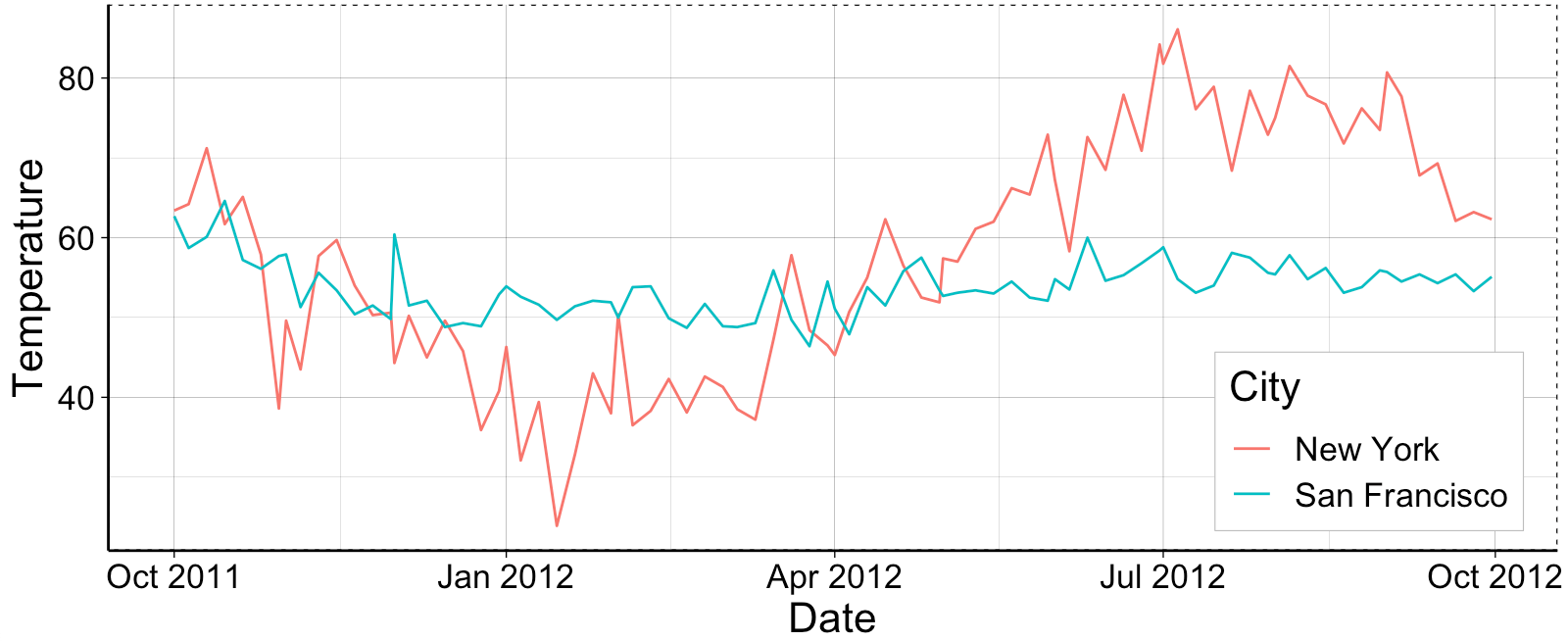}
 \caption{A line chart that shows temperature trends.}
 \label{fig:r-vis-1}
\end{subfigure}\\\medskip
\begin{subfigure}{0.98\linewidth}
\centering
\includegraphics[width=\textwidth]{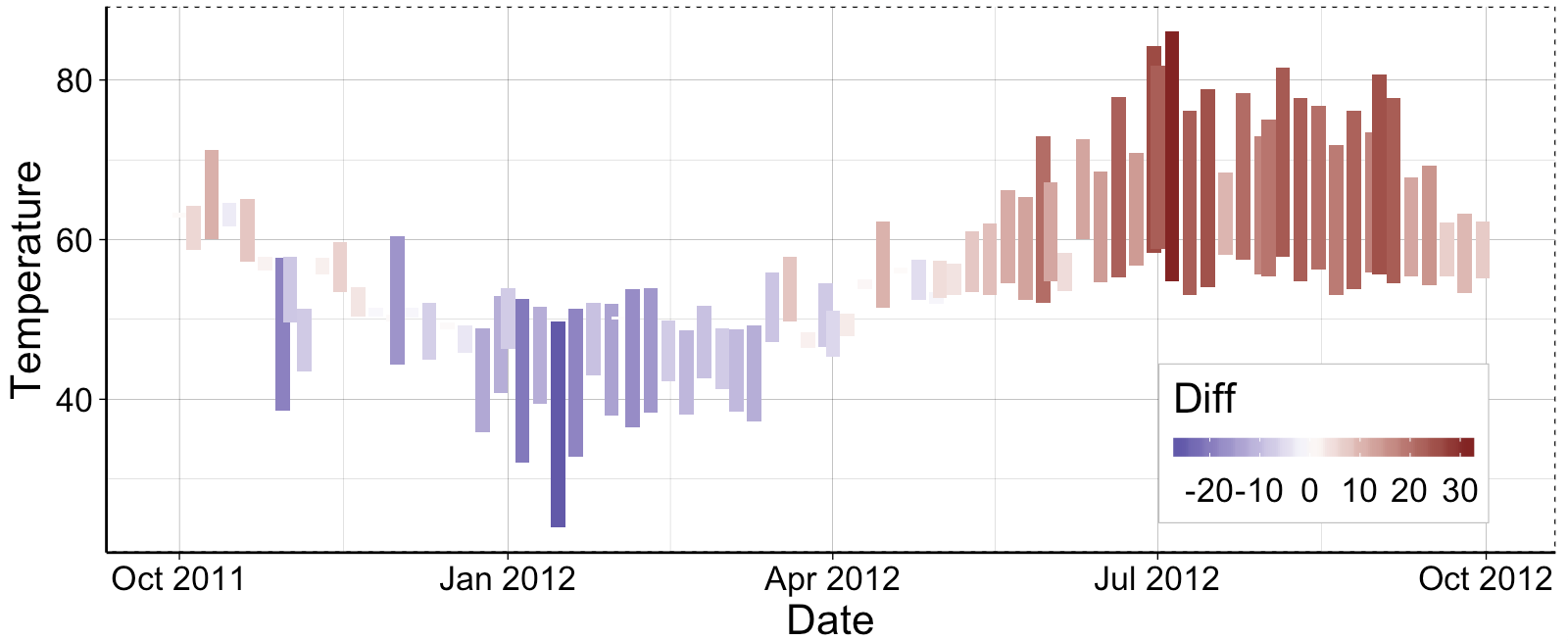}
 \caption{A bar chart that visualizes temperature difference.}
 \label{fig:r-vis-2}
\end{subfigure}
\vspace{-5pt}
 \caption{Two visualizations created in R that compare New York and San Francisco temperatures.}

\end{figure}

\begin{center}
\small
\vspace{5pt}
\centering
\begin{smpage}{0.9\linewidth}
\begin{minted}{R}
df1 <- pivot_longer(data = df, 
         cols = ("New York", "San Francisco"),
         names_to = "City", values_to = "Temperature")
\end{minted}
\end{smpage}~\\
\vspace{10pt}
\begin{smpage}{\linewidth}
\centering
\begin{tabular}{|c|c|c|}
\hline
Date & City & Temperature\\\hline
2011-10-01 & New York &  63.4\\
2011-10-01 & San Francisco & 62.7\\
... & ... & ... \\
2012-09-30 & San Francisco & 55.1\\\hline
\end{tabular}
\end{smpage}
\vspace{5pt}
\end{center}
After data transformation, Eunice specifies the visualization using the following script. The script maps \textsf{Date} to $x$-axis , \textsf{Temperature} to $y$-axis, and \textsf{City} to both color and group. It generates the visualization in \autoref{fig:r-vis-1}.

\begin{center}
\small
\vspace{3pt}
\begin{smpage}{0.96\linewidth}
\begin{minted}{R}
plot1 <- ggplot(data = df1) + 
         geom_line(aes(x = `Date`, y = `City`, 
            color= `Temperature`, group = `Temperature`)) 
\end{minted}
\end{smpage}
\vspace{5pt}
\end{center}

\noindent Eunice then proceeds to create bars on top of the first layer to visualize the temperature difference. Eunice first finds  the function \verb|geom_rect| from the library that supports floating bars. To visualize temperature difference, Eunice needs to specify positions of bars by mapping \textsf{Date} to $x_\mathit{min}$ and $x_\mathit{max}$ properties and mapping temperatures of the two cities to $y_\mathit{min}$ and $y_\mathit{max}$; she also needs to map the temperature difference between the two cities to $\mathit{color}$ to specify bar colors.
Since the original data does not contain a column for temperature difference, Eunice uses the \verb|mutate| function from tidyverse to transform the data. Using the following script, Eunice successfully creates the visualization in \autoref{fig:r-vis-2}.

\begin{center}
\small
\vspace{3pt}
\begin{smpage}{0.90\linewidth}
\begin{minted}{R}
df2 <- mutate(df, Diff = `New York` - `San Francisco`) 
plot2 <- ggplot(df2) + 
         geom_rect(aes(xmin = `Date`, xmax = `Date`, 
            ymin = `New York`, ymax = `San Francisco`, 
            fill = `Diff`))
\end{minted}
\end{smpage}
\vspace{5pt}
\end{center}

\noindent Finally, Eunice restructures the code to combine the two layers together using a concatenation operator. She also fine-tunes some parameters in ggplot2 to improve visualization aesthetics (e.g., modify titles of the axes and change line chart to a step chart), which generates the visualization that matches her design in \autoref{fig:nyc-sf-temp-diff}. 

Since Eunice is an experienced data analyst, she manages to go through these data transformation and visualization step and eventually generates the desired visualization. However, a less experienced data analyst, Amelia, finds the visualization task challenging.
\begin{itemize}[leftmargin=5.5mm]
\item First, Amelia is not familiar with the ggplot2 library, so she struggles in identifying the right functions to use. For example, it is difficult for her to distinguish between \verb|geom_path| and \verb|geom_line|, and \verb|geom_bar| or \verb|geom_rect|. She is also unfamiliar with how to compose multi-layered visualizations.
\item Second, due to her lack of experience with ggplot2, she finds it difficult to conceptualize the expected input layout because different functions and tasks require different data layouts.
\item Finally, due to her lack of experience with tidyverse, she needs to spend significantly more time in finding the right operators and implementing the desired transformation.
\end{itemize}

\subsection{User Experience in Falx}
\label{sec:falx-user-experience}
\begin{figure*}[t]
\centering
\includegraphics[width=0.98\linewidth]{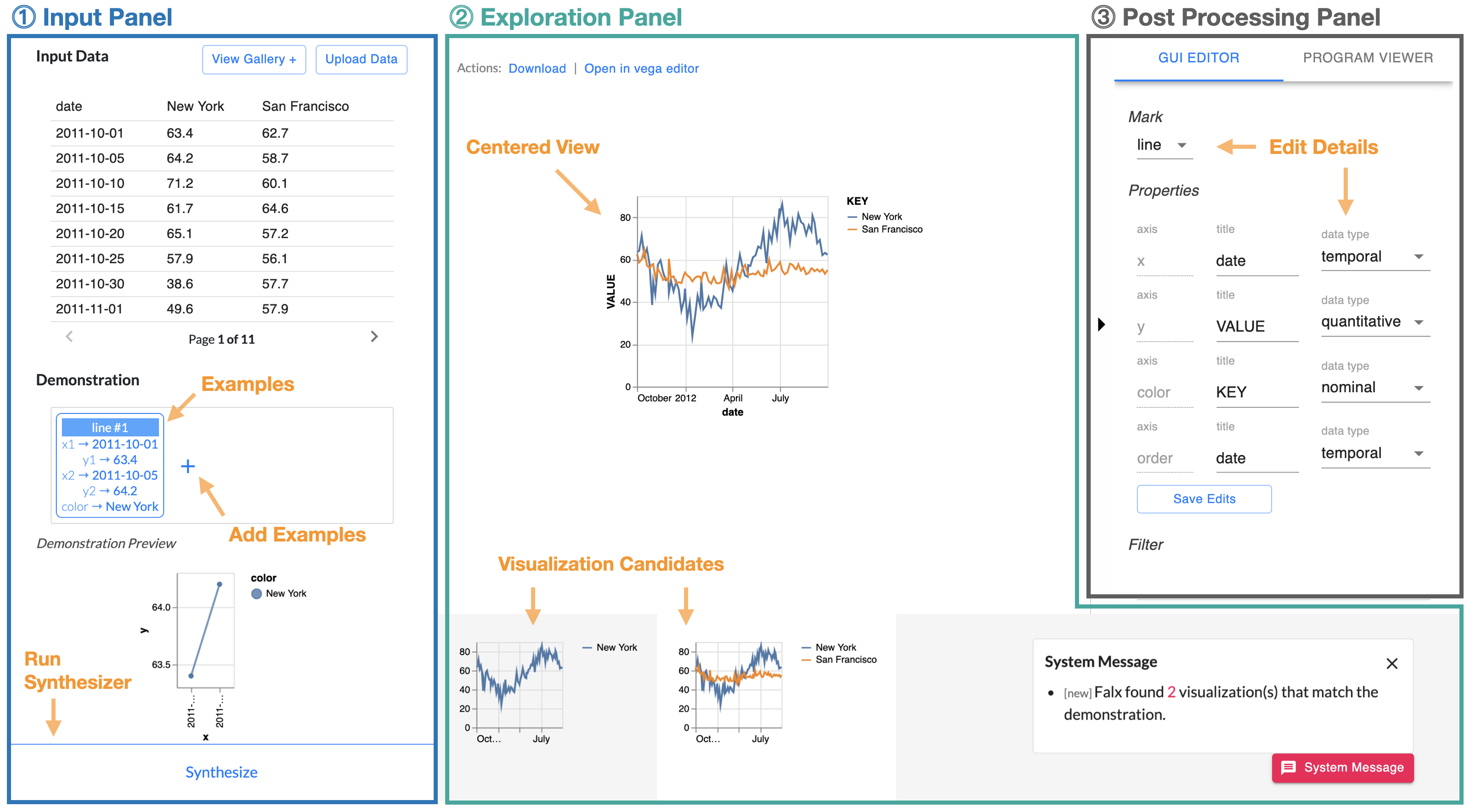}
\caption{Falx interface has three panels: (1) Data analysts import data and create examples in the input panel. (2) Analysts explore and examine synthesized visualizations in the exploration panel. (3) Analysts edit visualization details in the post processing panel.}
\label{fig:falx-interface}
\end{figure*}

Now we show how Amelia, a less experienced data analyst, uses Falx (\autoref{fig:falx-interface}) to create the same visualization.

\begin{figure*}[t]
\centering
\includegraphics[width=0.8\linewidth]{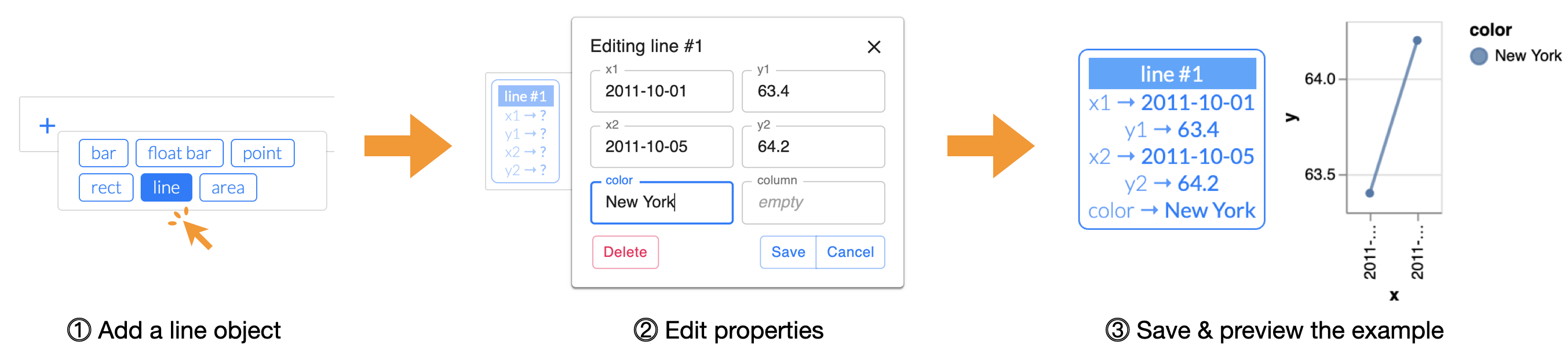}
\vspace{-5pt}
\caption{Amelia creates a line segment to demonstrate the visualization task.}
\label{fig:falx-create-example}
\end{figure*}

First, Amelia uploads the input data to Falx's input panel (\autoref{fig:falx-interface}-\textcircled{\small{1}}) and examines the input data displayed in a tabular view. Amelia decides to first visualize temperature trends of the two cities using a line chart. Amelia goes to the demonstration panel to demonstrate how the first two data points of New York temperatures will be visualized. To do so, Amelia first clicks the ``+'' icon in the interface and select a line element (\autoref{fig:falx-create-example}-\textcircled{\small{1}}), and Falx pops out an editor panel for Amelia to specify properties of this line element. Amelia clicks on values in the input table and copies the values to specify properties of the line element as follows (\autoref{fig:falx-create-example}-\textcircled{\small{2}}):
\begin{itemize}[leftmargin=5.5mm]
\item The line segment starts at the point with $x_1=\textsf{2011-10-01}$, $y_1=63.4$ (New York temperature on 2011-10-01)
\item The line ends at  $x_2=\text{2011-10-05}$, $y_2=64.2$ (New York temperature on 2011-10-05)
\item The color of the line is labeled as ``\textsf{New York}''
\end{itemize}

After saving the edits, Falx registers the example and provides a preview that visualizes the example line segment (\autoref{fig:falx-create-example}-\textcircled{\small{3}}) for Amelia to examine. Using this example, Amelia conveys the following visualization idea to Falx: ``I want a line chart over the input data that contains the demonstrated line segment''.
Amelia then presses the ``Synthesize'' button (in \autoref{fig:falx-interface}-\textcircled{\small 1}) to ask Falx to find the desired line chart. Internally, Falx first infers the visualization specification and then runs a data transformation synthesizer to transform the input data to match the visualization specification. After approximately four seconds, Falx finds two visualizations that match the example and displays them in the bottom of the exploration panel (\autoref{fig:falx-interface}-\textcircled{\small{2}}). Both visualizations contain the example line segment demonstrated by Amelia but they generalize the example differently: the first visualization only visualizes New York temperatures, while the second generalizes the color dimension to other columns in the input data as well, resulting in a visualization that also contains San Francisco temperatures. 

After briefly examining both candidates, Amelia finds the second visualization closer to the design in her mind, so she clicks the second visualization to enlarge it in the center view for a detailed check (\autoref{fig:falx-interface}-\textcircled{\small{2}} top). In the center view, Amelia hovers on the visualization to check details like values of different points in each line. After confirming the visualization matches her design, Amelia moves on to the second layer visualization, which should display temperature differences between the two cities using a series of bars. 

\begin{figure*}[h]
\centering
\includegraphics[width=0.85\linewidth]{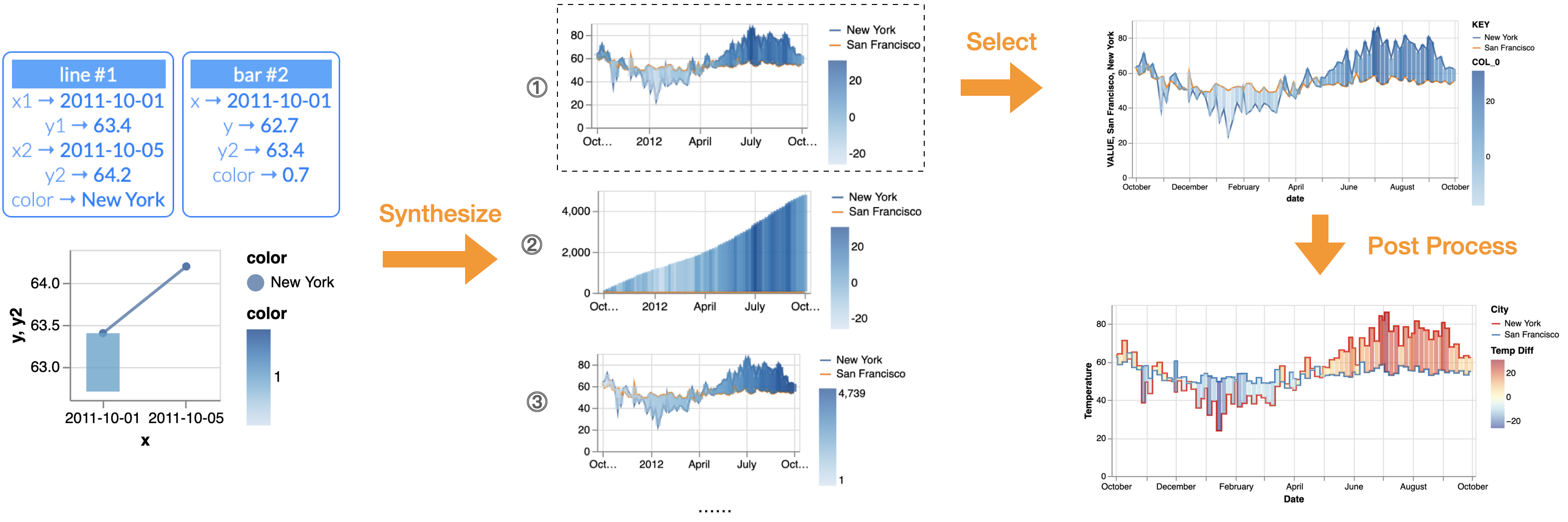}
\caption{Amelia's interaction with Falx to create the second layer visualization.}
\label{fig:falx-create-second-layer}
\end{figure*}

Next, 
Amelia creates an example bar to demonstrate how the temperature difference between the two cities on \textsf{2011-01-01} should be visualized (\autoref{fig:falx-create-second-layer} left): the bar is positioned at date \textsf{2011-10-01}, it starts at $62.7$ (San Francisco temperature), ends at $63.4$ (New York temperature), and its color shows the temperature difference of $0.7$ for that day. Amelia runs the synthesizer 
to find visualizations that contain both the example line and the example bar. This time, after 9 seconds, Falx finds 8 candidate visualizations that match the examples (\autoref{fig:falx-create-second-layer} middle). To decide which visualization to pick, Amelia can either (1) add a second example bar to demonstrate the temperature difference of the two cities on another date to help Falx resolve the ambiguity, or (2) navigate candidates in the exploration panel to examine them. Amelia decides to use the second approach again. She first rules out some obviously incorrect visualizations (e.g., visualization 2 in \autoref{fig:falx-create-second-layer} middle), then compares similar visualizations, and finally selects the first visualization to check it in detail. After some examination, she decides it matches her design and proceeds to post-process the visualization.

The post processing panel ( \autoref{fig:falx-interface}-\textcircled{\small{3}}) contains a GUI editor that allows Amelia to fine-tune visualization details and a program viewer for viewing and editing the synthesized program. Any changes made during the editing process are directly reflected on the center view panel (\autoref{fig:falx-interface}-\textcircled{\small{2}}) to provide immediate feedback. Using the post-processing panel, Amelia changes the line mark to step mark and modifies axis titles, which produces the visualization in \autoref{fig:falx-create-second-layer} right. Amelia is happy with this visualization and concludes the task. If Amelia wants to further customize the visualization (e.g., change color scheme, adjust bar spacing), she can directly edit the underlying Vega-Lite program.

In sum, Amelia creates the visualization by iterating through creating examples, exploring synthesized visualizations, and post processing. In this process, she benefits from the following design decisions behind  Falx:

\begin{itemize}[leftmargin=5.5mm]
\item First, while two visualization layers require different data transformations, Amelia does not need to worry about this, as the transformation task is delegated to the underlying synthesizer. In fact, even if the input data comes with a different layout, Amelia can still solve the problem with the same examples.
\item Second, Amelia specifies examples by choosing from a small set of visualization marks and specifying mappings from concrete data values to properties. This allows her to create visualizations without programming in the visualization grammar.
\item Third, instead of asking Amelia to read synthesized programs to disambiguate synthesis results, Falx provides an exploration interface that allows Amelia to explore and examine results in the visualization space.
\item Finally, Falx adopts a scalable synthesis algorithm to explore the exponential number of possible ways to transform and visualize the input data. Each synthesis  run takes between 3 and 20 seconds, which makes Amelia conformable at iterating between giving examples and exploring the generated visualizations.
\end{itemize}

\section{System Architecture}
In this section, we first provide a brief review of program synthesis and discuss the design and implementation of Falx, our end-to-end synthesis tool for automating data visualization tasks.

\subsection{Background: Program Synthesis}
In recent years, many program synthesis algorithms have been developed to automate challenging or repetitive tasks for end users by automatically generating programs from high-level specifications (e.g., demonstrations, input-output examples, natural language descriptions). For instance, programming-by-example (PBE) is a branch of program synthesis that aims to synthesize programs that satisfy input-output examples provided by the user, such tools been used for string processing~\cite{flashfill,singh2016}, tabular data transformation~\cite{morpheus,hades,scythe}, and program completion~\cite{sketch1,sketch2,slang,insynth,prospector}.

While there are different approaches to synthesize programs, one common method is to perform enumerative search over the space of programs by gradually expanding programs from a context-free grammar of some language~\cite{lambda2,DBLP:conf/cav/AlbarghouthiGK13,transit,hades}. In general, these search techniques traverse the program space according to some cost metric and return the candidate programs that satisfy the user-provided specification.
Here, the cost metric can be a model that measures simplicity of programs (e.g., based on number of expressions in the program)~\cite{lambda2} or a statistical models that estimate likelihood of the program being correct~\cite{deepcoder,slang}. 
To speed up the synthesis process, several recent methods use deduction rules to prune incorrect \emph{partial programs} early in the search process~\cite{lambda2,morpheus}. 
For instance, Morpheus~\cite{morpheus} uses predefined axioms of table operators to detect conflicts before the entire program is generated.

\subsection{Falx Synthesizer}
\begin{figure*}
\centering
\includegraphics[width=0.98\linewidth]{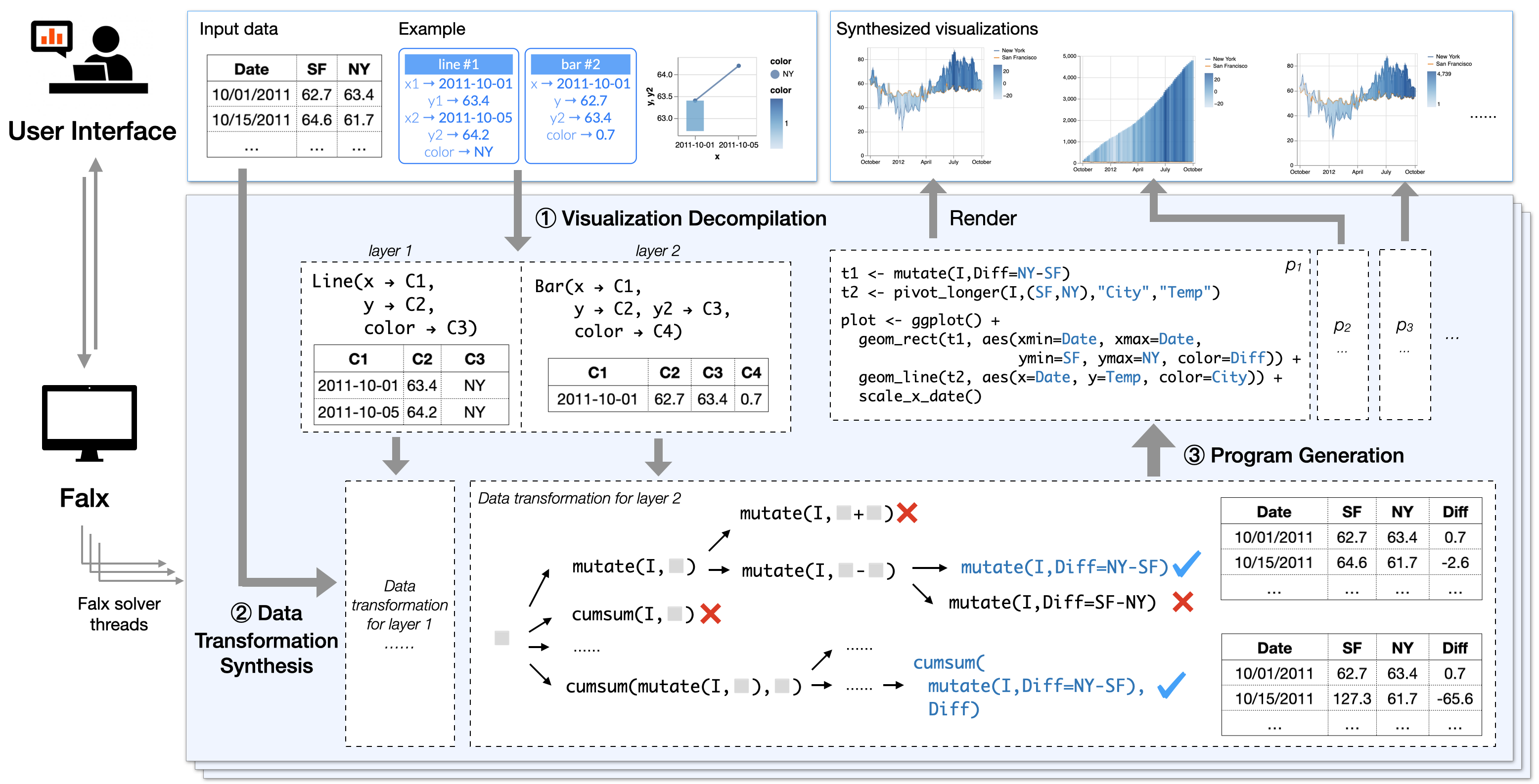}
\caption{The architecture of the Falx system. Each solver thread synthesizes visualizations that match user examples in three steps: (1) visualization decompilation, (2) data transformation synthesis, and (3) program generation.}
\label{fig:falx_system}
\end{figure*}

The architecture of Falx is shown in \autoref{fig:falx_system}. To use Falx, a data analyst first provides an input table and creates examples to demonstrate the visualization idea. Once the analyst hits the ``synthesize'' button, the Falx interface sends the input and examples to the Falx server. Given an input data and an example visualization (in the form of a set of geometric objects), Falx synthesizes pairs of candidate data transformation and visualization programs such that the resulting visualization contains all geometric objects in the visualization example.

To synthesize visualizations consistent with examples from the user, Falx spawns multiple solver threads to solve the synthesis problem in parallel. In each solver thread, Falx first runs a \emph{visualization decompiler} (step 1) to decompile the example visualization into a visualization program and an example table, such that applying the program on the example table yields the example visualization provided by the user. Then, Falx calls the \emph{data transformation synthesizer} (step 2) to infer programs that can transform the input data to a table that contains the example table generated in step 1. Finally, for each candidate data transformation result, Falx generates a \emph{candidate visualization} (step 3) by combining the transformed data with the visualization program synthesized in step 1 and compiling them to Vega-Lite or R scripts for rendering. Synthesized visualizations from all threads are collected and displayed in Falx's exploration panel for the analyst to inspect. In what follows, we elaborate on the details of each step using the same running example in Section~\ref{sec:usage-scenario}.

\subsubsection*{Step1: Visualization Decompilation}

Internally, Falx represents visualizations as a simplified visualization grammar similar to ggplot2 and Vega-Lite. In this grammar, a visualization is defined by (1) graphical marks (line, bar, rectangle, point, area), (2) encodings that map  data fields to visual channels ($x$, $y$, size, color, shape, column, row), and (3) layers, which specify how basic charts are combined into compositional charts. Since Falx only uses this grammar as an intermediate language to capture visualization semantics, visualization details (e.g., scale types) are intentionally omitted. 
Falx goes through the following three steps to decompile a visualization.
\begin{itemize}[leftmargin=5.5mm]
\item Falx first infers visualization layers from the user example. In particular, Falx partitions examples provided by the user into groups based on their geometric types and properties, and creates one visualization layer for each group. Each layer corresponds to a simple chart of a particular type (e.g., scatter plot, line chart).
\item Then, for each layer, Falx creates one basic visualization and an example table. The example table contains the same number of columns as the number of visual channels in this layer (derived from properties of geometric objects), and the visualization is specified as encodings that map columns in the example table to visual channels.
\item Finally, for each example table, Falx fills the table with values from the example geometric objects. 
\end{itemize}

\begin{example}
As shown in \autoref{fig:falx_system}-\textcircled{\small{1}}, given the two visual elements provided by the user, Falx infers that the desired visualization should be a multi-layer chart that is composed by a line chart in layer 1 and a bar chart in layer 2 and decompiles the two layers independently. For example, for the second layer, Falx generates a bar chart program $\mathsf{Bar}\{x \mapsto \mathsf{C1}, y \mapsto \mathsf{C2}, y_2 \mapsto \mathsf{C3}, \textit{color} \mapsto \mathsf{C4}\}$ with an example table $T=[(\textit{2011-10-01}, 62.7, 63.4, 0.7)]$ where $T$ represents the desired \emph{output table} that should be the result of the data transformation process. Column names $\mathsf{C1}, ..., \mathsf{C4}$ in the bar chart program correspond to names of the four columns in Table $T$.
\end{example}

\subsubsection*{Step 2: Data Transformation Synthesis}
After decompiling the examples into the visualization program and example tables $T$, together with the original input table $T_\mathsf{in}$ provided by the user, Falx reduces the visualization synthesis task into a data transformation synthesis task~\cite{morpheus,scythe,wang2019visualization}. For each example table $T$, the data transformation synthesizer aims to synthesize a transformation program $P_{t}$ that can transform the input table into a table that contains the example table, i.e., $T \subseteq P_{t}(T_\mathsf{in})$.
Falx supports various types of transformation operators commonly used in the \texttt{tidyverse} library to handle different layouts of the input from the user (\autoref{fig:data-transform-operators}).

\begin{figure*}[ht]
\begin{tabular}{|l|l|l|}
\hline
Type & Operator & Description \\\hline
\multirow{2}{*}{Reshaping} & \verb|pivot_longer| & Pivot data from wide to long format\\
& \verb|pivot_wider| & Pivot data from long to wide format\\\hline
\multirow{2}{*}{Filtering} & \verb|select| & Project the table on selected columns\\
& \verb|filter| & Filter table rows with a predicate \\\hline
\multirow{3}{*}{Aggregation} & \verb|group| & Partition the table into  groups based on values in selected columns\\
& \verb|summarise| & For every group, aggregate values in a column with an aggregator\\
& \verb|cumsum| & Calculate cumulative sum on a column for each group\\\hline
\multirow{3}{*}{Computation} & \verb|mutate| & Arithmetic computation on selected columns\\
& \verb|separate| & String split on a column \\
& \verb|unite| & Combine two string columns into one with string concatenation\\\hline
\end{tabular}
\caption{Data transformation operators supported in Falx. For clarity, we omit the parameters of each operator.}
\label{fig:data-transform-operators}
\end{figure*}

The data transformation synthesizer uses an efficient algorithm to search for  programs that are compositions of operators in \autoref{fig:data-transform-operators} satisfying the requirement $T \subseteq P_{t}(T_\mathsf{in})$. Falx starts the search process by constructing sketches of transformation programs (i.e., programs whose arguments are not filled) and then iteratively expands the search tree and fills arguments in these partial programs. To maintain efficiency in this combinatorial search process,  Falx uses deduction to prune infeasible partial programs as early as possible (as used in prior work~\cite{morpheus,scythe,wang2019visualization}). The deduction engine analyzes properties of partial programs using abstract interpretation~\cite{cousot1977abstract} and prunes programs whose analysis results are inconsistent with the example output. Since each partial program corresponds to several dozens of concrete programs, the deduction engine can dramatically prune the search space.

When the search algorithm encounters a concrete program (i.e., with all arguments are filled) that is consistent with the example output, Falx adds the program to the candidate pool. The search procedure terminates either when the designated search space is exhaustively visited or when the given search time budget is reached. All synthesized program candidates are sent to the post-processor to generate visualizations.

\begin{example}
\autoref{fig:falx_system}-\textcircled{\small{2}} shows the data transformation synthesis process for the second visualization layer (the bar chart) generated in step \textcircled{\small{1}}. Given the original input table $I$ (with three columns \textsf{Date}, \textsf{SF}, and \textsf{NY}) the output table $T$ (with four columns \textsf{C1}, \textsf{C2}, \textsf{C3}, and \textsf{C4}) generated in the last step, Falx aims to transform $I$ into a table that contains the example table $T$. Starting from an empty program, Falx iteratively expands the unfilled arguments (represented as holes ``$\Box$'') in the partial programs to traverse the search space. 
When Falx encounters a partial program $\mathsf{cumsum}(I, \Box)$, Falx abstractly analyzes it and concludes that it is infeasible because \textsf{cumsum} cannot transform an input table with three columns into an output table with four columns.
Falx the expands the feasible partial programs (e.g., $\mathsf{mutate}(I, \Box)$) and collects concrete programs that are consistent with the objective (e.g., $\mathsf{mutate}(I, \mathsf{Diff}=\mathsf{NY}-\mathsf{SF})$).
\end{example}

\paragraph{Optimization.} We made several optimizations on top of existing synthesis algorithms~\cite{morpheus,wang2019visualization} to reduce Falx's time to respond. First, the major overhead in synthesis is the cost of analyzing partial programs using abstract interpretation, as it often requires running expensive operators like aggregation and pivoting on big tables. To reduce this overhead, Falx memoizes abstract interpretation results for partial programs to allow reusing then whenever possible. 

Second, instead of aiming to find only on or a few candidate programs that match user inputs like prior algorithms, Falx expects to find as many different programs as possible that satisfy the examples to ensure the correct visualization is included. To ensure diverse outputs, different Falx solver threads start with different initial program sketches to search for different portions of the search space in parallel. To improve responsiveness, Falx sets different timeouts for different threads to allow faster threads to respond to the user while other threads are searching for more complex transformations. In our implementation, we run 2 solver threads in parallel, we set one thread with 5 seconds timeout and another with 20 seconds timeout based on our perception of how long an analyst would be willing to wait as well as the typical time Falx takes to finish traversing different parts of the search space.

\subsubsection*{Step 3: Processing Synthesized Visualizations}

As the final step in visualization synthesis, Falx generates visualizations by combining the visualization program generated in step 1 with table transformation programs generated in step 2.

Concretely, for each data transformation program, Falx applies the table transformation program on the input data to obtain a transformed output and unifies the output table schema with the schema in the visualization program, since the visualization program was filled with placeholder column names \textsf{C1}, \textsf{C2}, ..., etc. Falx then instantiates other visualization details (e.g., scale type, axis domain, etc.) omitted in the visualization grammar and compiles the visualization program into a Vega-Lite (or R) script through syntax-directed translation. For example, in \autoref{fig:falx_system}-\textcircled{\small{3}}, Falx generates an R script that both transforms the input and specifies the visualization. Furthermore, Falx notices that the values on the $x$-axis are dates instead of strings, so it changes the $x$-axis scale to a temporal scale using the function ``\verb|scale_x_date()|''.

After compilation, the post-processor removes semantically duplicate visualizations (i.e., visualizations with different specifications but with the same content and detail). Finally, Falx groups and ranks the visualizations based on the complexity of the programs (numbers of expressions). In this way, similar visualizations are grouped together to make comparison easier in the exploration process, and the complexity ranking allows users to explore visualizations constructed from easier transformation programs first before jumping into complex ones. These visualizations are sent to the user interface for rendering to allow user exploration.

\section{User Study}\label{sec:user-study}

To understand Falx’s benefits and limitations and to examine how analysts might adopt synthesis-based visualization tools, we conduct a between-subjects evaluation centered on the following questions:
\begin{itemize}[leftmargin=5.5mm]
\item Does Falx improve user efficiency in creating visualizations compared to a baseline tool?
\item How does Falx change the visualization authoring process for different data analysts?
\item What strategies do data analysts use to visualize data in Falx?
\end{itemize}

\subsection{Participants}\label{sec:participants}
We recruited two groups participants for the study: 16 participants (10 M, 5 F, 1 Unknown, Ages 23-51) for the Falx study, and another 17 participants (12 M, 4 F, Ages 19-60) for the baseline tool study (the R programming language). In the recruiting process, we screened participants by their ability to read a sample visualization. For the baseline group, we additionally required that all participants have experience with R (specifically ggplot2 and tidyverse libraries) for data visualization. 

Participants reported their experience in data visualization authoring based on the number of visualizations they created in the past 6 months using any tools. For the Falx study group, there were 6 participants experienced with some visualization tools (created >10 visualizations), 8 with moderate experience with visualization tools (created 1-10 visualizations), and 2 participants with zero experience in creating visualizations in the past. For the baseline group, there were 8 experienced participants (create >10 visualizations) and 9 participants with moderate experience (created 1-10 visualizations).

\begin{figure*}[t]
\centering
     \begin{subfigure}[b]{0.27\textwidth}
         \centering
         \includegraphics[width=\textwidth]{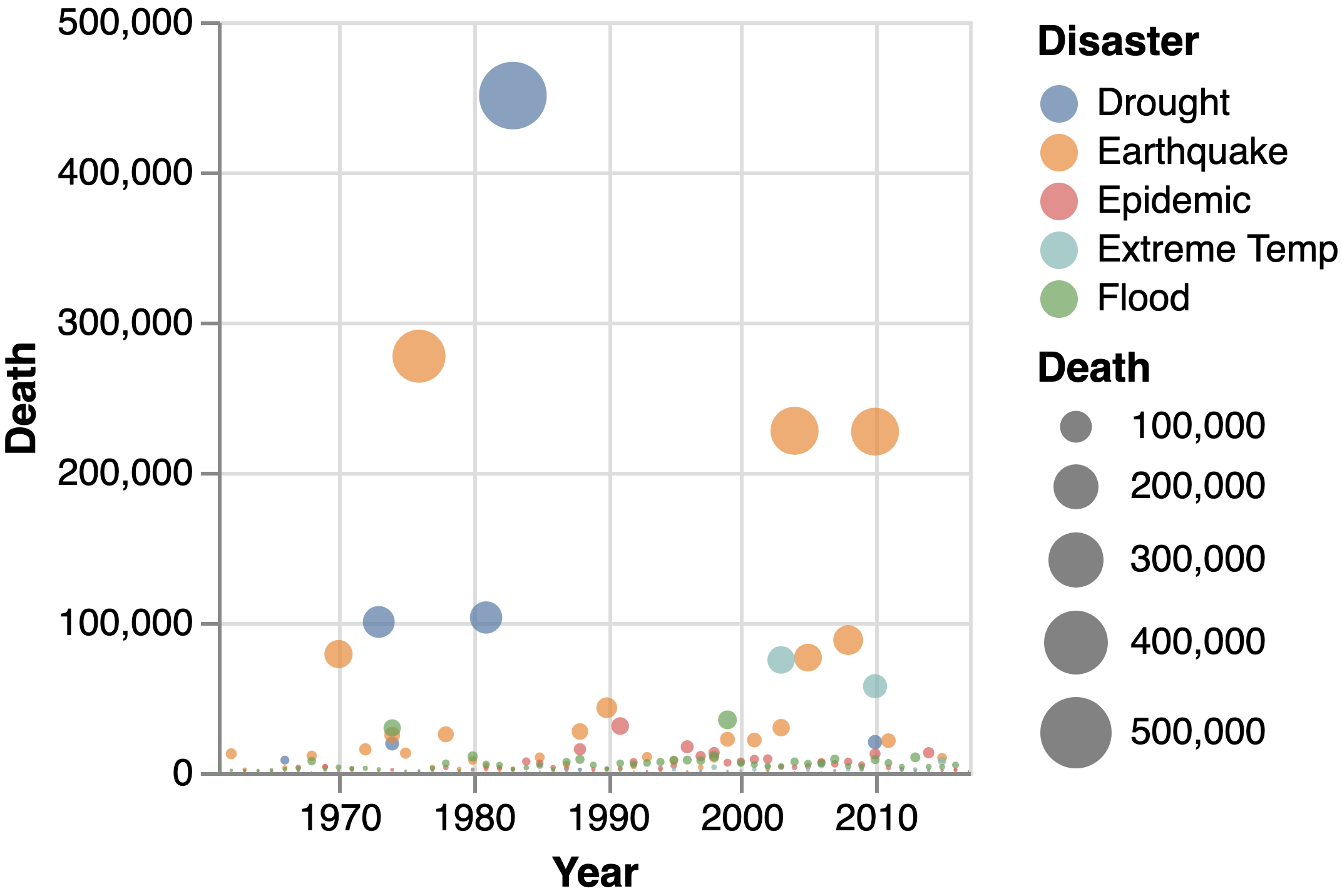}
         \caption{Disaster impact}
     \end{subfigure}
     ~
     \begin{subfigure}[b]{0.18\textwidth}
         \centering
         \includegraphics[width=\textwidth]{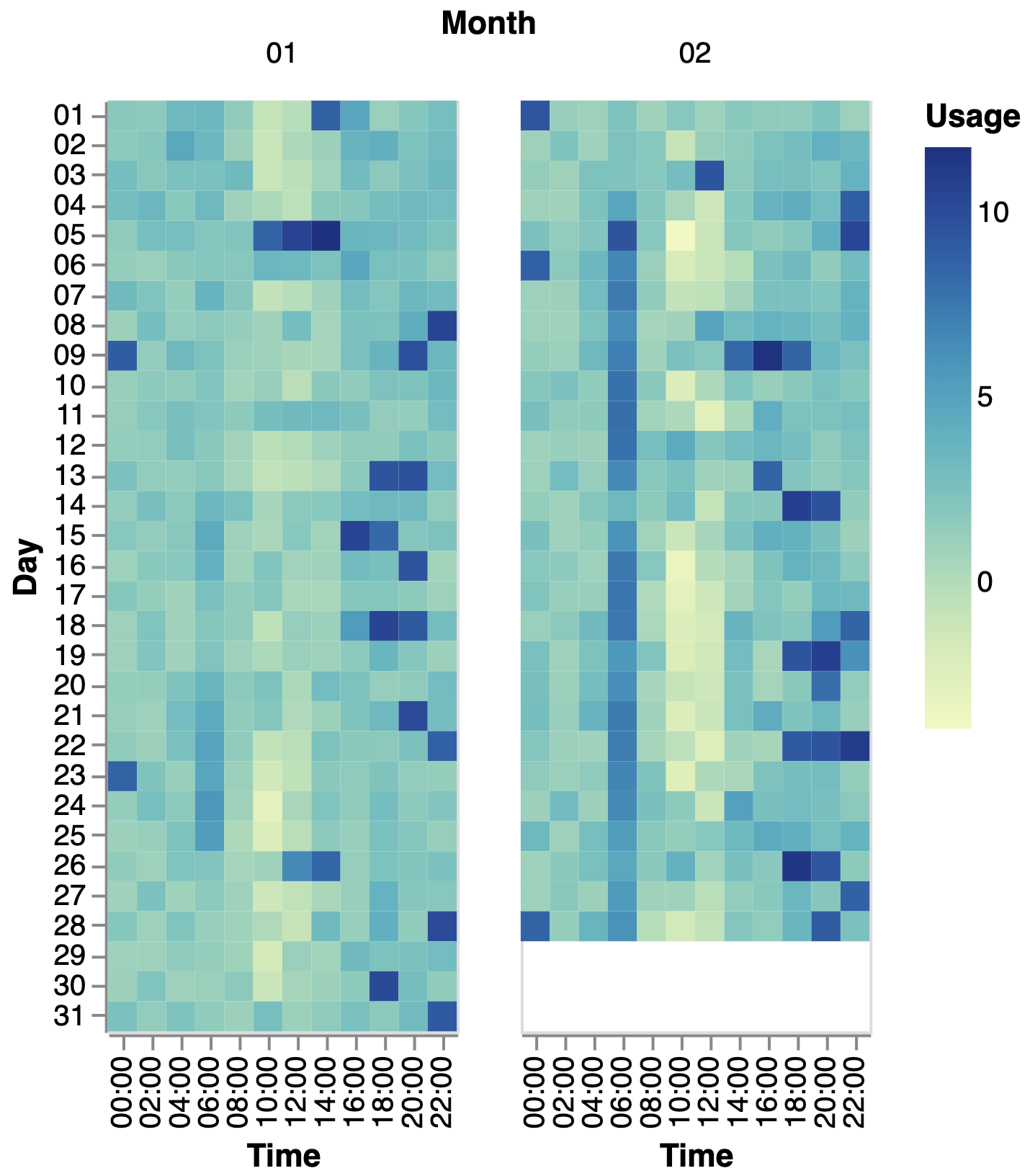}
         \caption{Electric usage}
     \end{subfigure}
     ~
     \begin{subfigure}[b]{0.27\textwidth}
         \centering
         \includegraphics[width=\textwidth]{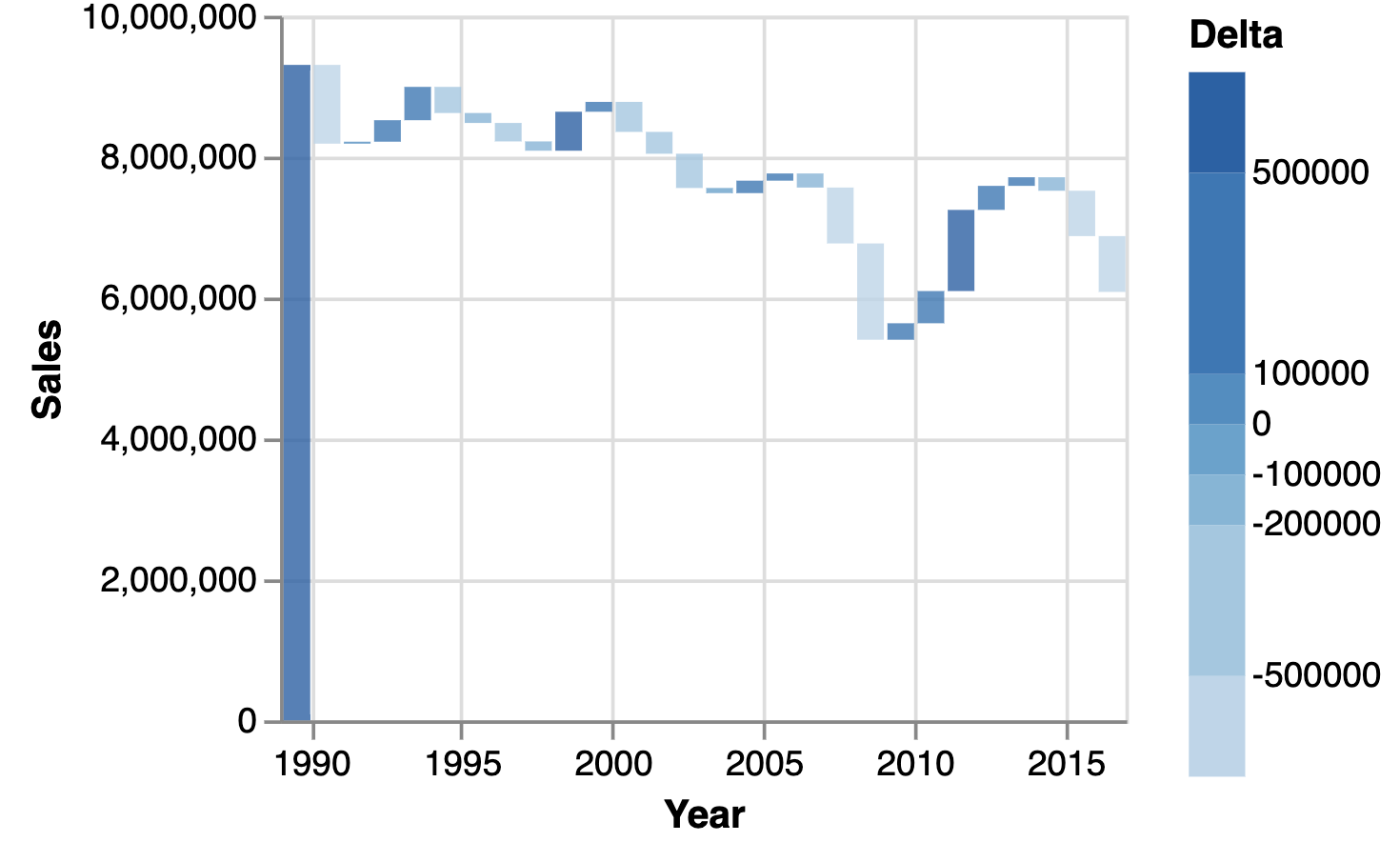}
         \caption{Car sales}
     \end{subfigure}
     ~
     \begin{subfigure}[b]{0.27\textwidth}
         \centering
         \includegraphics[width=\textwidth]{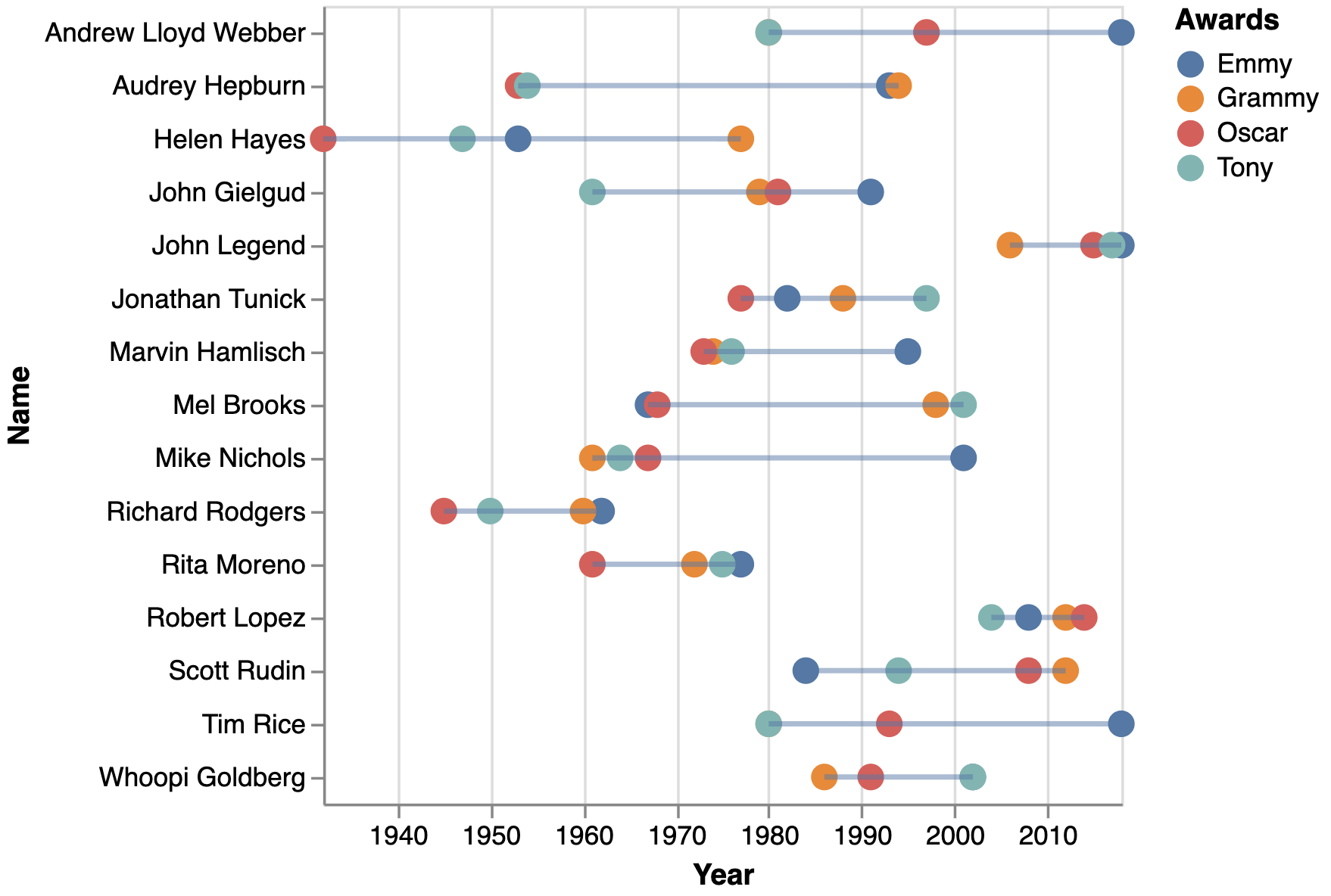}
         \caption{Movie awards winners}
     \end{subfigure}
     
\caption{Study tasks.}
\label{fig:study-task}
\end{figure*}

\subsection{Procedure}
\label{sec:study-procedure}
Each participant was asked to complete four visualization tasks, where the Falx study group completed the task using Falx and the baseline group used R to complete the task. We chose R as the baseline tool due to its popularity among data analysts and its ability to support both data transformations and visualizations in the same context, where many other visualization tools requires users to process data and specify visualizations in different contexts.

To better examine the use of Falx, participants in the Falx group first completed a 20-minute tutorial together with a warm-up task with a sample solution (creating a grouped line chart to visualize sea ice level change in the past 20 years). After the tutorial, participants were asked to solve four visualization tasks. For R participants, we also provided the same warm-up task with a sample solution to allow users to get familiar with the environment and the data loading process, so that participants could focus on solving the visualization tasks. During the user study, participants were allowed to refer to any resource on the Internet including documentations and QA forums. We collected screen and audio recordings while participants completed tasks. We then interviewed them after all tasks were completed to reflect on their visualization process and strategies.

To conduct our user study, we developed four different visualization scenarios (\autoref{fig:study-task}): 
\begin{enumerate}[label=(\alph*), leftmargin=5.5mm]
    \item \emph{Disaster Impact}: A scatter plot that visualizes the number of people died from five disasters in the last century.
    \item \emph{Electric Usage}: A faceted heat map for hourly electric usage in each day during the first two months of 2019.
    \item \emph{Car Sales}: A waterfall chart for the number of cars sold in a year. Each bar starts at the sales value in the previous month and ends at the sales values in the month, and its color gradient reflects the increase/decrease compared to the last month.
    \item \emph{Movie Awards}: A layered line/scatter plot for visualizing winners of all four prestigious movie awards. For each celebrity, there are four points showing years these awards were earned and a line showing the time span for the celebrity to win all four awards.
\end{enumerate}
For each visualization task, we provided as input a table that can be directly imported into the tools. We also explicitly described visualization designs to the participants in text so that participants could focus on implementation. Finally, we asked participants that they do not need to optimize the design --- a task was considered correctly solved as long as the semantics of the visualization created by the participant matched the example solution regardless of the process and details. In this study, we did not restrict the time participants could spend on each task, but we provided users the option of quitting a task after spending more than 20 minutes without success.
Thus,  participants could complete each task with one of three outcomes: (1) submit a correct solution, (2) submit a wrong solution, or (3) give up after trying for at least 20 minutes. 

We interviewed each participant after they finished all four tasks. For both Falx and baseline groups, we interviewed participants about (1) challenges they encountered while solving the tasks and their solutions, (2) common errors they made and how they fixed them, (3) their confidence about the solutions they submitted and what checks they performed to ensure correctness, and (4) what additional resources they used during the study and how they helped. We additionally asked participants in the Falx group to reflect on their visualization authoring process and interviewed them about (1) strategies adopted when creating examples to demonstrate the visualization task, (2) strategies adopted to explore the synthesized visualizations, and (3) their prior visualization experience and how Falx could potentially fit in their routine work.

The total session was less than 2 hours for all participants. To address learning effects or other carryover effects, we counterbalanced the tasks using a Latin square. We performed our analysis using mixed effect models, treating participants as a random effect and modeling tool, tasks, and experience level as fixed effects.

\subsection{Task Completion}

\autoref{tab:user-success-rate} shows the percentage of participants that correctly finished each task.  Falx participants generally had higher completion rates in all tasks. We observed a statistically significant difference in the completion rate in the car sales visualization ($p<0.05$); others were not significant. Among nine failed tasks by Falx users, seven were due to incorrect solutions and, in two cases, participants quit the task after 20 minutes. Among 20 failed cases in the R study group, there were 9 incorrect solutions ans 11 cases where participants quit after 20 minutes.

\begin{figure}[t]
\centering
\begin{tabular}{|c|cc|cc|}
\hline
\multirow{2}{*}{Task} &  \multicolumn{2}{|c|}{R $(N=17)$}  & \multicolumn{2}{|c|}{Falx $(N=16)$} \\
 & $n$ & \% & $n$ & \% \\\hline
Disaster Impact & 16 & 94.1\% & 14 & 87.5\%\\
Electric Usage  & 13 & 75.6\% & 14 & 87.5\%\\
Car Sales       & 5  & 29.4\% & 11 & 68.8\%\\
Movie Awards    & 14 & 82.4\% & 16 & 100\%\\\hline
\end{tabular}
\caption{The number and percentage of participants correctly finished each study task.}
\label{tab:user-success-rate}
\end{figure}

\begin{figure}[t]
\includegraphics[width=0.47\textwidth]{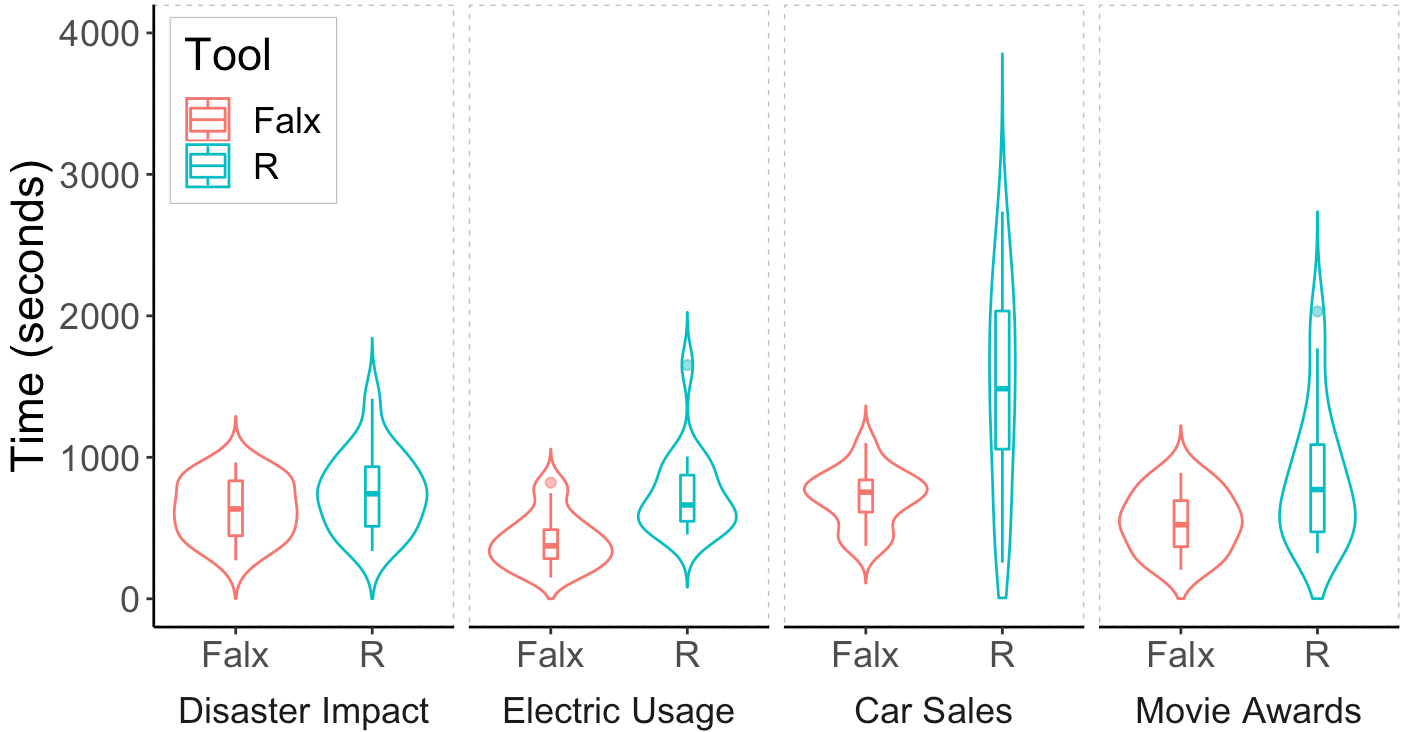}
\caption{Violin plot showing the amount of time participants spent on each task for both Falx and R study groups.}
\label{fig:user-efficiency}
\end{figure}

\autoref{fig:user-efficiency} shows task completion time in Falx. Using Wilcoxon rank sum test with Holm’s sequential Bonferroni procedure for $p$ value correction, we observed a significant improvement in user efficiency for car sales visualization ($t_\mathsf{Falx}=715\pm202s$, $t_\mathsf{R}=1473\pm743 s$, $\mathit{\mu}_\mathsf{R} - \mathit{\mu}_\mathsf{Falx} = 758s$, $p < 0.01$)~\footnote{We use $t_\mathsf{Falx}$ and $t_\mathsf{R}$ to show the mean and standard deviation of time participants in Falx and R groups spent on each task. We use $\mu_\mathsf{R}-\mu_\mathsf{Falx}$ to represent the difference of the mean time between the two groups.} and electric usage visualization ($t_\mathsf{Falx}=411\pm192 s$, $t_\mathsf{R}=740\pm297s$, $\mathit{\mu}_\mathsf{R} - \mathit{\mu}_\mathsf{Falx} = 329s,p < 0.001$). While Falx participants were also generally faster in the other two tasks, there was no significant difference for the  movie industry celebrity visualization ($t_\mathsf{Falx}=544\pm215 s$, $t_\mathsf{R}=861\pm490 s$, $\mathit{\mu}_\mathsf{R} - \mathit{\mu}_\mathsf{Falx} = 323s, p=0.07$) or the disaster impact visualization ($t_\mathsf{Falx}=638\pm209 s$, $M_\mathsf{R}=754\pm279s$, $\mathit{\mu}_\mathsf{R} - \mathit{\mu}_\mathsf{Falx} = 116s, p=0.23$). Participants from the R study group noted that the key reasons for failing on the car sales visualization task was the difficulty of finding the correct API (for waterfall chart) together with the complex transformation behind it (which required calculating a cumulative sum). Falx users also noted they found the car sales visualization difficult due to unfamiliarity with the visualization type. On the other hand, R users reported that the movie awards visualization and the disasters impact visualization were relatively easier since they expected the same pivot operator to transform the input, which is commonly encountered by R users, and the visualization types were relatively standard (line chart and scatter plot).

We found no significant interaction between user experience level (defined in \autoref{sec:participants}) and task completion time ($p=1$ for all tasks in both study groups using Wilcoxon rank sum test with Holm’s sequential Bonferroni correction).

\subsection{Task Experience}
\label{sec:qualitative-feedback}

In this section, we describe qualitative feedback from participants in both groups about general (non-Falx related) visualization challenges both during the study and in their daily work, and how Falx can help with solving some of these challenges. We leave discussions of Falx-specific visualization challenges to \autoref{sec:visualization-process-falx}.

As described in \autoref{sec:study-procedure}, we conducted a semi-structured interview for participants of both groups about visualization challenges they encountered both in the study and in their daily work, and how some of these challenges are typically overcome. To analyze this data, two of the researchers collaboratively conducted a qualitative inductive content analysis on the interviewer’s notes, with a sensitizing concept of \emph{visualization challenges and solutions}. In this process, two researchers independently labeled interview notes and then collaboratively discussed and compared high level labels to resolve disagreements in the initial codes.

\subsubsection{Finding the right visualization function} The first challenge frequently mentioned by participants was discovering or recalling the correct visualization function. In the R study group, 14 out of 17 participants described this challenge, especially for the car sales task that most participants failed on. Some participants noted that the difficulty came from both finding the right term to search and distinguishing similar candidate functions. For example, participant R14~\footnote{We use R1-R17 to denote participants from the R study group and F1-F16 to denote participants from the Falx group.} noted that
{\it ``I wasn't aware that $\mathit{geom\_rect()}$ would be more helpful than $\mathit{geom\_bar()}$. One thing that made it more challenging was the fact that this kind of bar chart has no proper name. I tried searching `non-contiguous bar charts in R', but I didn't get many useful results.''}. These challenges are also common in compositional charts: R10 noted {\it ``creating the line with the dots is something I never did before so didn't know how to achieve it''}. 
To address these challenges, participants noted that online example galleries and forums are {\it ``essential to their work''} (R1). Besides, two participants had {\it ``an internal file -- R code dictionary''} (R7) and {\it ``a collection of some own code snippets''} (R1) to reduce search effort. 

Falx group participants also described that they faced similar challenges of finding right functions in their daily work and Falx could help address them. For example, F1 mentioned:
{\it ``Falx can generate something that you cannot easily do. For example, the multi-layered visualization for the movie dataset would be very difficult to do in Excel or Google doc, you may need to specify some formula to specify relationship between two layers.''} Participant F11 mentioned that Falx helped with complex tasks because {\it ``It allows you to start by creating a relatively simple visualization in the beginning, which is good, then it allows you to build more complex stuff on top of it which is also helpful.''}

\subsubsection{Data transformation} Data transformation was another frequently mentioned challenge, including both conceptualizing the expected data layout and implementing the transformation. For example, 
R17 mentioned {\it ``it [the car sales task] also seems to require some extra aggregation to get the starting and ending value for each rectangle to be drawn, which makes it even more difficult.''} About implementation, R9 said that {\it ``the vocabulary of the tidyverse is critical for trying to do what you want to do, otherwise it is all impossible to achieve.''}, and R14 mentioned that {\it``I had an idea of what I needed to do, but I wasn't able to search the right things on Google to arrive at a useful code snippet for it.''}

Participants from the Falx group mentioned similar issues in their work routine. For example, {\it ``Tableau won’t do data preparation and you need to manually put them together''} (F7), {\it ``pivoting table is already something at an intermediate level in Tableau and many people cannot use it''} (F2). Due to lack of skill of preparing data programmatically, some participants would do it manually. For example, {\it ``if I need to pivot data, I do it manually -- e.g., just copy the data to a blank area [in Excel] and pivot it''} (F8).
Participants appreciated that Falx automatically handled data transformations. Participant F5 mentioned {\it ``I like the fact that it [Falx] solves the data transformation and visual encoding. I’m pretty familiar with visual encoding so it is fine  when the data is in the right shape. But I find transforming data annoying.''} Participant F15 mentioned {\it ``I didn’t think about data format at all in the process''}. F7 mentioned {\it ``Tableau won’t do data preparation because you need to manually put them together and drag drop them for you. Falx is pretty automated on this.''}

\subsubsection{Learning to create expressive visualizations} Due to the inherent challenge in visualization and data transformation in these tools, participants mentioned many of existing tools had a learning barrier for new users. For example, F4 mentioned that {\it ``the learning curve is pretty steep (Tableau), and we spent a lot of time learning these tools''}. On the other hand, while Falx was a new visualization tool, most users found it easy to learn, despite some users requiring some time in the beginning to get used to {\it ``the paradigm shift from my normal understanding''} (F6). For example, participant F4 mentioned that {\it ``the ramp up time [for Falx] is pretty short and it’s pretty easy to use.''}, and F6 mentioned that {\it ``anyone with basic Excel knowledge should be able to use Falx''}.

\subsection{Visualization Strategies in Falx}
\label{sec:visualization-process-falx}

Since Falx is a new tool for data visualization, besides understanding its ability to address existing visualization challenges, we also investigated how participants used Falx to solve visualization tasks. We conducted an inductive content analysis on the interviewer’s notes about Falx experience similar to that in \autoref{sec:qualitative-feedback}. In this section, we discuss observations about participants' visualization process in Falx and their indications for future synthesizer-based visualization tool design.

\subsubsection{Strategies for creating examples}\label{sec:strategy-create-example} Data analysts initiate  interactions with Falx by creating examples. As a synthesis-powered visualization tool, poorly constructed examples can be highly ambiguous and lead to long running time and a large number of visualization candidates. Also, while users can carefully create multiple examples to increase Falx's performance, it requires more effort. Falx users identified the following strategies to create examples effectively:
\begin{itemize}[leftmargin=5.5mm]
	\item \emph{Sketching visualizations before demonstration:} Three participants mentioned that sketching the visualization design on paper helped them understand geometry of the visualization, and it helped them creating better examples. For example, participant F13 mentioned {\it ``I sketch out first to get a general understanding of what the visualization would look like, and then use that to drop points.''}. 
	\item \emph{Selecting representative data points to demonstrate:} Seven participants mentioned that they considered using  {\it ``representative points''} (F7) when creating demonstrations in order to reduce ambiguity to Falx. For example, participant F1 mentioned that {\it ``[In the disaster impact task], I chose a cause that contains non-zero value in that year, because it’s a unique value that can avoid confusion of the tool''}. 
	\item \emph{Start from a few examples, add more later if necessary:} Eight participants mentioned that they {\it ``tried to shoot for minimum input''} (F6) for simplicity. In this way, they can {\it ``run the tool to see what it returns''} (F1) before spending more effort on examples, and they would {``add more to help narrow it down if there are many visualizations pop up''} (F9). Additionally, participant F11 noted that {\it ``It’s easy to add multiple elements to mess up with the demonstration. A small number of elements make it easier to go back and fix''}.
	\item \emph{Start with multiple examples to minimize interaction iterations:} Instead of starting from minimal inputs, 6 participants preferred to create more examples in the beginning to {\it``avoid ambiguity''} (F2). They remarked that {\it ``it doesn't take that much time to add data points''} (P8) and multiple examples can {\it ``avoid having to wait and choosing from multiple solutions''} (F8).
\end{itemize}

During the process of creating and revising examples, seven participants found the demo preview panel useful since it allowed them to {\it `` understand more about how a certain layout would look like''} (F11) and it {\it ``helps put me on the right track of solving the task.''} (F13). However, nine participants said they did not find it helpful because they {\it ``don’t know if it tells enough to help understand anything [about synthesis results]''} (F7); they preferred to {\it ``just click synthesis to get the result since synthesis is pretty fast''} (F14).

Some challenges participants encountered in creating examples included (1) unfamiliarity with terms in Falx (e.g., F4 mentioned {\it ```size' is a term that I’m not familiar with.''}) and (2) not getting used to demonstrate visualization ideas using values (e.g., F6 mentioned {\it `'I was struggling with the paradigm shift about when to use values and when to use table headers''}). In general, the fast response time of Falx enabled participants to get over these challenges through trial and error (e.g., F1 mentioned {\it ``If there is anything wrong, I’ll go back and do edits on the points.''}), and they {\it ``get faster in later tasks once understand the difference''} (F6). In future, Falx could adopt a mixed-initiative interface~\cite{kandel2011wrangler} to improve experience for new users.  In addition, we observed that many participants felt like they were interacting  with an intelligent tool (e.g., F13 mentioned {\it ``the tool is quite good at learning from what I demonstrated''}) and they were willing to provide more informative inputs (e.g., F16 {\it ``tried to write the expression because I don’t know how Falx would do computation''}). In future, Falx could take advantage of this to support more complex visualization tasks by synthesizing programs from users more informative inputs  besides examples (e.g., formulas that describe how certain values in the examples are derived from the input). 

\subsubsection{Strategies for exploring synthesis results} After creating examples to demonstrate the visualization task, users interact with Falx to explore the synthesized visualizations and identify the desired solution. Prior work~\cite{DBLP:journals/aim/Lau09,DBLP:conf/uist/MayerSGLMPSZG15} has shown that a main barrier for adoption of synthesis-based programming tools is that users have difficulty understanding and trusting synthesized solutions, especially when there are many solutions consistent with the user demonstration. 

We discovered from the interview that many participants shared the following similar 4-step process to select the desired visualization from synthesized visualizations by investigating visualization from coarse to fine:

\begin{itemize}[leftmargin=5.5mm]
	\item \emph{Step 1: Check against the high-level picture.} First, participants noted that it was easy to quickly exclude many visualizations that are obviously far from the desired visualization. For example, {\it ``having too many options is a bit overwhelming, but just keeping in mind what the result you look like can help narrow down the solution''} (F11).
	\item \emph{Step 2: Check axes and invariants.} After excluding the obviously wrong solutions, participants often investigate domains and ranges of each axis to further refine synthesis results. For example, {\it ``I first looked at color labels, I noticed they tend to be wrong in wrong visualizations -- e.g., some charts only contain 2 labels instead of 4''} (F16).
	\item \emph{Step 3: Compare similar visualizations.} Then, participants investigated similar visualizations to find their difference. For example, {\it ``In the electric case, there is one mistake [in a candidate visualization] with 2019 showing up on y axis, it’s small and not obvious. But then, I was able to tell the difference by comparing the two visualizations directly, and notice that year showed up in the 'hour' field''} (F2).
	\item \emph{Step 4: Inspect visualization detail.} Finally, participants {\it ``check carefully about the values to make sure they are correct''} (F5). An example of such detailed checking is to check values in the chart against known values in the input data: {\it ``if there is a specific value that I know is correct -- for example, in the last example (disasters), I knew the total death for 1961 was, then I hover over the output to check if the value is correct ''} (F6). 
\end{itemize}
After these steps, participants were confident about the result. In fact, while participants mentioned that their confidence about solutions could be negatively affected by unfamiliarity of visualization types (e.g., F9 mentioned {\it `` I don’t do much heatmap so I’m less confident''}), they mentioned that the checking process can raise their confidence about the chosen solution. For example, participants got more confident after {\it ``comparing them [candidates] with my sketch''} (F6), {\it ``looking at solutions and finding their difference''} (F14), or {\it ``checking details''} (F2). They further noted that in many cases, {\it ``it’s almost impossible for Falx to get it wrong because these values are all pretty unique''} (F14). In general, participants found the exploration panel {\it ``quite useful''} because it {\it ``allows to choose the best visualization out of that''} (F7).

In sum, Falx's exploration panel allowed users to directly inspect solutions in the visualization space following a coarse-to-fine process, which helped them to disambiguate solutions and trust the chosen results. In the future, Falx's interface could be improved to augment users' exploring strategies. For example, Falx could directly summarize the differences among the synthesized visualizations to allow users to make comparisons easier. Also, Falx's center view panel could support displaying traces that show how properties of each geometric object are derived from the input, which could make the synthesis process more transparent and make checking details easier.

\subsection{Workflow Implications}

Finally, participants reflected on how Falx might fit into their workflow. For example, F13 mentioned {\it ``I’ll absolutely use this if this is a product. Even as it is now I’ll use it''}. Participants found several scenarios that Falx can be helpful.
\begin{itemize}[leftmargin=5.5mm]
    \item Create visualizations for discussions and presentations. For example, F1 noted that {\it ``visualizations generated by Falx can meet standards of presentation slides''} and {\it ``Falx can generate something that you cannot easily do in Excel''}. 
    \item Prototyping complex analysis. For example, F16 mentioned {\it ``Falx is very useful in the prototyping stage because it’s very fast to use.''} F7 further noted that they can {\it ``take a sample to visualize and then extend to the full visualization''} using Falx for analyzing big datasets. 
    \item Benefit non-experienced users. Six participants mentioned that Falx can be {\it ``more beneficial to new users that cannot create charts''} (F2). Also, Falx can be {\it ``a good teaching tool to help people understand data''} (F7).
    \item Reduce team collaboration effort. Participant F11 described that visualization readers were often different from visualization creators in their team, and modifying visualizations required team efforts. F11 mentioned that Falx could help with it: {\it ``a person presents me with a visualization, but I want to view something differently. Instead of getting back to the person to re-do it, I can probably just use Falx, which would be more efficient.''} 
\end{itemize}

However, several participants also mentioned Falx may not fit well to their current workflow when they need {\it ``very high standard visualizations''} (F1) that requires extensive customization. Another limitation of the current version of Falx  is the lack of {\it ``deep integration with other tools''} (F1), e.g., database for handling big datasets and data cleaning tools for {\it ``handling null / dirty data''} (F4). But in general, participants thought that Falx would be helpful when used  in the right scenarios and {\it ``would be pretty interesting to try Falx in some of these tasks''} (F5).
\section{Related Work}
Falx builds on top of prior research on grammar based visualization tools, data transformation tools, program synthesis algorithms and automated visualization design systems.

\paragraph{Grammar-based Visualization}
Following the initial publication of the Grammar of Graphics~\cite{wilkinson2012grammar}, high level grammars~\cite{DBLP:journals/tvcg/SatyanarayanMWH17,wickham2011ggplot2,DBLP:conf/kdd/StolteTH02} for data visualizations have grown increasingly popular as a way of succinctly specifying visualization designs. In contrast to low level visualization languages like Protovis~\cite{bostock2009protovis}, D3~\cite{DBLP:journals/tvcg/HeerB10}, and Vega~\cite{DBLP:journals/tvcg/SatyanarayanRHH16} that are designed for creating highly-customizable explanatory visualizations, these high level grammars aim to enable analysts to rapidly construct expressive graphics in exploratory analysis. 
For example, ggplot2~\cite{wickham2010layered,wickham2011ggplot2} and Vega-Lite~\cite{DBLP:journals/tvcg/SatyanarayanMWH17} are two visualization grammars that allow users to specify visualizations using visual encodings. In both tools, low level visualization details are handled by default parameters unless users want  customization. Tableau~\cite{DBLP:conf/kdd/StolteTH02} adopts a graphical interface approach to enable users to rapidly create views to explore multidimensional database. In Tableau, users drag-and-drop data variables onto visual encoding ``shelves'', which are later translated into a high-level grammar similar to ggplot2. These tools expect the input data layout to match the design such that (1) each row corresponds to a graphical object, and (2) each column can be mapped to a visual channel. 
In practice, the mismatch between the design and the input data layout is common, which raises a barrier for creating visualizations~\cite{gatto2015making,wongsuphasawat2019goals}.

Falx formalizes visualizations in the same way, and synthesized programs are compiled to ggplot2 or Vega-Lite for rendering. Falx's user interface also inherits the expressiveness and simplicity of Grammar of Graphics design, by allowing users to create examples of visual encodings to demonstrate visualization ideas. The main difference is that Falx relaxes the constraints on input data layout and allows users to use layout-independent examples to demonstrate visualization ideas. Falx then automatically infers the visualization spec and synthesizes data transformations to match the data with the design from the examples, which saves users' construction efforts. 

\paragraph{Data Transformation Tools}
The need to prepare data for statistical analysis and visualization has led to the development of many tools for data transformation~\cite{raman2001potter,wickham2014tidy,kandel2011wrangler,drosos2020wrex}. Since different analysis objective requires different layout, users need to frequently transform data throughout the analysis process~\cite{wickham2014tidy,kandel2011research,wongsuphasawat2019goals}.
Potter’s Wheel~\cite{raman2001potter} is a graphical interface that allows users to interactively choose transformation operators and inspect transformation outputs. Wrangler~\cite{kandel2011wrangler} is a mixed initiative data transformation tool which can suggest transformations based on the input data. Tidyverse~\cite{wickham2014tidy} is a data transformation library in R, which allows users to interleave data transformation code, analysis code and visualization code in the same environment to reduce the effort of context switch. Several synthesis-powered data transformation tools~\cite{scythe,morpheus,drosos2020wrex,polozov2015flashmeta,barowy2015flashrelate} have been proposed to help automate data transformation. For example, Prose~\cite{polozov2015flashmeta} includes several programming-by-example tools that automatically synthesize programs for data cleaning and transformation from input-output examples. Morpheus~\cite{morpheus} and Scythe~\cite{scythe} are two specialized data transformation synthesizer with better scalability and expressiveness.

Falx inherits the transformation language design in tidyverse~\cite{wickham2014tidy}, and Falx is a realization of prior program synthesis algorithms~\cite{morpheus,wang2019visualization} as an interactive system for visualization authoring. Falx's main difference from  automated data transformation tools is the unification of the visualization task and transformation tasks. In this way, Falx users do not need to conceptualize expected data layout or frequently switch between visualization and data transformation tools. The unification also enables Falx users to easily explore synthesis results in the visualization space as opposed to program space, which is considered challenging~\cite{DBLP:conf/uist/MayerSGLMPSZG15}. Besides data layout transformation, many data preparation tools also support data cleaning (e.g., handling missing data or invalid data)~\cite{DBLP:journals/pacmpl/WangDS17}, data normalization (collecting non-relational data into relation format)~\cite{barowy2015flashrelate}, and string formatting~\cite{drosos2020wrex,flashfill,zhanginteractive}. Falx currently does not support directly visualizing dirty or non-relational data. In the future, Falx could work with these tools to further automate visualization process.

\paragraph{Visualization Automation}

Automated visualization  tools~\cite{DBLP:journals/tvcg/MoritzWNLSHH19,hu2019viznet,saket2016visualization} have been proposed to help data analysts to explore the visualization design space. Draco~\cite{DBLP:journals/tvcg/MoritzWNLSHH19} and Dziban~\cite{lin2020dziban} use constraint logic approaches to model design knowledge, and they can recommend visualization designs from partial specifications. VizNet~\cite{hu2019viznet} uses a deep neural network trained from visualization corpus to suggest designs. Voyager~\cite{wongsuphasawat2015voyager} combines recommendation and exploration for mixed-initiative design exploration. VisExemplar~\cite{saket2016visualization} allows users to demonstrate changes in the visualization layout to explore alternative visualizations designs. Falx is complementary to these design automation tools. Falx allows users to implement visualization designs they have in mind without data layout constraints, while design automation tools helps users to explore visualization designs from a fixed data layout. A combination of the two approaches could potentially help users to explore a larger visualizations design space without data layout constraints.

\paragraph{User Interaction with Program Synthesizers} In general, program synthesizers can be categorized into exploration tools and implementation tools. Synthesis-based exploration tools aim to generate a large number of solutions from users' weak constraints to aid users to explore the search space~\cite{swearngin2020scout,DBLP:journals/tvcg/MoritzWNLSHH19}.
For example, Scout~\cite{swearngin2020scout} is a synthesis-based exploration tool to discover mobile layout ideas. In these tools, users interact with an exploration interface to navigate and save interesting solutions. Implementation tools~\cite{scythe,morpheus,flashfill,polozov2015flashmeta,zhanginteractive,drosos2020wrex}, instead, aim to synthesize programs to help solve a concrete task (e.g., implement a design that a user already have in mind). In these tools, the main interaction objective is to help users to disambiguate spurious programs that happen to be consistent with the user specification but are incorrect for the full task~\cite{DBLP:conf/uist/MayerSGLMPSZG15}. To solve this challenge, Wrex~\cite{drosos2020wrex} generates readable programs for users to inspect and edit; Regae~\cite{zhanginteractive} and FlashProg~\cite{DBLP:conf/uist/MayerSGLMPSZG15} interactively ask users disambiguating questions to refine synthesis results; PUMICE~\cite{DBLP:conf/uist/LiRJSMM19} lets users collaborate with the agent to recursively resolve any ambiguities or vagueness through conversations and demonstrations.

Falx is an implementation tool for data visualization. Falx's contribution to the user interaction model is that Falx brings the exploration design (from exploration tools) to address the disambiguation and trust challenges in implementation tools. Allowing users to explore and examine synthesized programs in the visualization space reduces the barrier for user interaction (e.g., users do not need to be familiar with underlying programs to disambiguate~\cite{DBLP:conf/uist/MayerSGLMPSZG15}) and increases users' confidence about solutions.

\paragraph{Tools for More Expressive Visualizations} Besides tools for standard visualization authoring, many visualization tools have been proposed to let designers create more expressive visualizations. Examples of these tools are Data illustrator~\cite{DBLP:conf/chi/LiuTWDDGKS18}, Lyra~\cite{DBLP:journals/cgf/SatyanarayanH14}, Charticular~\cite{DBLP:journals/tvcg/RenLB19}, Data-driven Guides~\cite{DBLP:journals/tvcg/KimSLDLPP17}, and StructGraphics~\cite{tsandilas2020structgraphics}. Besides high-level design layout (e.g., x,y ,column) and standard mark properties (e.g., color, shape), these tools let users customize marks to create more expressive glyphs (e.g., compound marks, parametric marks). These tools expect users to prepare data into a tidy format to start with, but they support rich visualization designs. Falx, in comparison, supports standard visualization designs but automates data transformation.

Several design reconstruction tools (e.g., VbD~\cite{saket2016visualization}, Liger~\cite{DBLP:journals/corr/abs-1907-08345}, iVolVER~\cite{DBLP:conf/chi/MendezNV16}) are proposed to let designers create expressive visualization by destructing and reconstructing existing visualization designs. Using these tools, users can transform existing visualizations to new ones by demonstrating desired design changes. Functionally, these tools are design exploration tools that take as input a visualization design and produce a new visualization design. They differ from Falx because Falx takes data as input and maps it to a visualization design for initial design authoring.

There are opportunities to combine Falx with these tools for better visualization authoring. Falx can work with expressive designs tools to support authoring complex visualizations from non-tidy data: users can first design customized marks using example data values, and the tool would automatically synthesize binding between data and these fine-grained mark properties from these examples. Falx can also work with design reconstruction tools to allow users to first use Falx to create initial design from data, and then subsequently interactively explore new designs by transforming the initial design. 
\section{Discussion}

We have presented Falx, a novel synthesis-powered visualization authoring tool that let users demonstrate a visualization design using examples of visual encodings and then receive suggestions for visualization designs. Our goal was to create a system that does not require users to manually specify the visualization or worry about data transformations, thereby improving user efficiency and reducing the learning burden on novice analysts. Our study found that Falx often achieved these goals: Falx users were able to effectively adopt Falx to solve visualization tasks that they could otherwise cannot solve, and in some cases, they do so more quickly. We next discuss some implications of this work in guiding future research.

\paragraph{Data Layout-Flexible Visualization Exploration} Besides visualization authoring, combining Falx with data exploration tools like GraphScape~\cite{2017-graphscape} Voyager~\cite{wongsuphasawat2015voyager}, GraphScape~\cite{2017-graphscape}, VbD~\cite{saket2016visualization} and Dziban~\cite{lin2020dziban} might enable new design exploration tools that allow users to discover both new relations from the dataset and new designs to visualize the them. Using existing design exploration tools, users can explore diverse visualization designs from an input data; but since existing tools generates designs that are specific to the input data layout, the design space that can be explored is limited. Integrating Falx with these design exploration tools could enable novel design exploration tools that can assist users to explore design space without being constrained by data layouts. For example, in an anchored design exploration scenario~\cite{lin2020dziban,2017-graphscape,saket2016visualization}, users can demonstrate data layout changes alongside design changes using this new tool to incrementally discover data insights from a larger design space. Similarly, Falx might also work with visualization recommendation engines~\cite{DBLP:journals/tvcg/MoritzWNLSHH19,hu2019viznet} to find better designs for the dataset based on initial visualizations created by users using examples to suggest data layout independent designs.

\paragraph{Visualization Learning} As we discovered from our study, users often describe existing programming tools as ``flexible, powerful'' but ``having a steep learning curve.'' Falx can fill in this gap by helping data analysts to learn to create visualizations. Since Falx does not require its users to have programming expertise, new users can learn visualization and data transformation concepts using Falx by first creating visualization using demonstrations and then inspecting synthesized programs. For example, Falx could generate readable code like Wrex~\cite{drosos2020wrex} for users to learn to use visualization APIs, enabling them to access the flexibility and power of code.

\paragraph{Bootstrapping Complex Data Analysis} Falx currently focuses on inexperienced data analysts, but it could also potentially benefit experienced data analysts by bootstrapping complex data analysis tasks. For example, data analysts could first create visualizations in Falx and then build complex analyses by iteratively editing synthesized programs. To achieve this goal, Falx needs more transparency and better integration with programming environments. For example, Falx could expose synthesized programs during the synthesis process and allow users to steer the synthesis process to better disambiguate results. Falx could also be integrated into programming environments like mage~\cite{DBLP:conf/uist/KeryRHMWP20}, Wrex~\cite{drosos2020wrex} or Sketch-n-Sketch~\cite{DBLP:conf/uist/HempelLC19} to make program editing easier.

\medskip

All of these possibilities, as well as prior work applying program synthesis to design (e.g., \cite{DBLP:journals/tvcg/MoritzWNLSHH19,swearngin2020scout}), suggest a promising future for augmenting design work with synthesis-based techniques. We hope Falx provides one exemplar for how to adapt core techniques in synthesis into powerful interactive tools that empower human creativity.

\begin{acks}
This work has been supported in part by 
the NSF Grants ACI OAC--1535191, FMitF CCF-1918027, OIA-1936731, IIS-1546083, IIS-1955488, IIS-2027575, CCF-1723352,
the Intel and NSF joint research center for Computer Assisted Programming for Heterogeneous Architectures (CAPA NSF CCF-1723352),
Department of Energy award DE-SC0016260,
the CONIX Research Center, one of six centers in JUMP, a Semiconductor Research Corporation (SRC) program sponsored by DARPA CMU 1042741-394324 AM01,
a grant from DARPA, FA8750--16--2--0032, 
as well as gifts from Adobe, Facebook, Google, Intel, VMWare and Qualcomm. 
We would also like to thank anonymous reviewers for their insightful feedback on paper revision. 
\end{acks}

\bibliographystyle{ACM-Reference-Format}
\bibliography{falx}

\end{document}